  \documentclass[final,3p,times,numbers,sort&compress]{elsarticle}   
\usepackage{hyperref}
\hypersetup{colorlinks,allcolors=black}
\usepackage{layouts}
\usepackage{amsmath,amssymb,stmaryrd,amstext,amsthm}
\usepackage{mathtools}
\usepackage{bm}	
\usepackage{caption}
\usepackage{subcaption}
\usepackage{siunitx}
\usepackage{booktabs}
\usepackage{multirow, graphicx}
\usepackage[english]{cleveref}
\usepackage[T1]{fontenc}

\usepackage{lscape}
\usepackage{pdflscape}
\usepackage[table, svgnames, dvipsnames]{xcolor}
\usepackage{longtable}
\usepackage{tabulary}
\usepackage{rotating,array,ragged2e,caption}
\usepackage{tabularx}
\usepackage{subcaption, nicefrac, lineno, float}

\usepackage{array}
\usepackage{makecell, cellspace, caption}

\newcommand{\blue}{\color{black}} %Blue NavyBlue

\newcommand{\black}{\color{black}}

\newcommand{\volsym}{\rlap{\kern.08em--}V} % Volume symbol

\def\tsc#1{\csdef{#1}{\textsc{\lowercase{#1}}\xspace}}
\tsc{WGM}
\tsc{QE}
\tsc{EP}
\tsc{PMS}
\tsc{BEC}
\tsc{DE}
\journal{International Journal of Heat and Fluid Flow} 

\begin{document}

\begin{frontmatter}



\title{Progressive augmentation of Reynolds stress tensor models for secondary flow prediction by computational fluid dynamics driven surrogate optimisation}

% Progressive augmentation of RANS models for secondary flow prediction by CFD-driven surrogate optimisation

\author[inst1, inst2]{Mario Javier Rinc\'on \corref{cor2}}
%\ead {mjrp@mpe.au.dk}
\author[inst1]{Ali Amarloo \corref{cor2}}
%\ead {amarloo@mpe.au.dk}
\author[inst2]{Martino Reclari}
%\ead {marr@kasmtrup.com}
\author[inst3]{Xiang I. A. Yang}
%\ead {xzy48@psu.edu}
\author[inst1]{Mahdi Abkar\corref{cor1}}
\ead{abkar@mpe.au.dk}

\cortext[cor1]{Corresponding author}
\cortext[cor2]{M. J. Rinc\'on and A. Amarloo contributed equally to this study.}

\address[inst1]{Department of Mechanical and Production Engineering, Aarhus University, 8200 Aarhus N, Denmark}
\address[inst2]{Quality \& Sustainability Department, Kamstrup A/S, 8660 Skanderborg, Denmark}
\address[inst3]{Department of Mechanical Engineering, Pennsylvania State University, State College, PA, 16802, USA}

\begin{abstract}
Generalisability and the consistency of the \textit{a posteriori} results are the most critical points of view regarding data-driven turbulence models. This study presents a progressive improvement of turbulence models using simulation-driven Bayesian optimisation with Kriging surrogates where the optimisation of the models is achieved by a multi-objective approach based on duct flow quantities. We aim for the augmentation of secondary-flow prediction capability in the linear eddy-viscosity model $k-\omega$ SST without violating its original performance on canonical cases e.g. channel flow. \blue Progressively data-augmented explicit algebraic Reynolds stress models (PDA-EARSMs) \black for $k-\omega$ SST are obtained enabling the prediction of secondary flows that the standard model fails to predict. The new models are tested on channel flow cases guaranteeing that they preserve the successful performance of the original $k-\omega$ SST model. Subsequently, numerical verification is performed for various test cases. Regarding the generalisability of the new models, results of unseen test cases demonstrate a significant improvement in the prediction of secondary flows and streamwise velocity. These results highlight the potential of the progressive approach to enhance the performance of data-driven turbulence models for fluid flow simulation while preserving the robustness and stability of the solver. %The new models developed are publicly available at \href{https://github.com/AUfluids/PA2DEARSM}{https://github.com/AUfluids/PA2DEARSM}.
\end{abstract}

\begin{highlights}
\item Two models with different levels of complexity have been developed to augment the $k-\omega$ SST turbulence model for the prediction of secondary flows.
\item The progressive approach involves improving the linear eddy-viscosity model's capacity to predict secondary flows, without compromising its successful performance in canonical flows.
\item The new \blue progressively data-augmented explicit algebraic Reynolds stress models (PDA-EARSMs) \black are able to predict secondary flows in contrast to common linear eddy-viscosity models.
\item CFD-driven surrogate optimisation has proven to be a robust method to obtain enhanced models that do not compromise the solver's stability.
\item The surrogate and Pareto-front solutions obtained are validated numerically and the best models obtained are verified against a series of canonical cases where their performance has been quantified.
\end{highlights}


\end{frontmatter}

% \linenumbers

\section{Introduction}

Reynolds-averaged Navier-Stokes (RANS) equations are widely preferred over high-fidelity methods like direct numerical simulation (DNS) and large-eddy simulation (LES) for the industrial applications of computational fluid dynamics (CFD) due to their robustness and computational speed. In RANS, the physics of turbulence is predicted by a Reynolds stress tensor (RST) model; hence, the results obtained are dependent on the performance of these model predictions. Despite the popularity of RANS simulations, the common empirical models have been found to have shortcomings \citep{slotnick2014cfd}, particularly in capturing Prandtl's second kind of secondary flow \citep{nikitin2021prandtl}. This limitation is accentuated in the most commonly used RANS turbulence models (e.g. two-equation models based on the Boussinesq assumption like $k-\varepsilon$ and $k-\omega$) due to their inability to predict complex turbulence anisotropy.

Although the development of reliable RANS turbulence models has remained stagnant for decades \cite{xiao2019quantification}, the recent advances in data-driven techniques have motivated a new wave of studies aiming at improving the performance of RANS turbulence models \cite{duraisamy2019turbulence}.
The majority of these studies have used available high-fidelity data of the RST values to train a model and improve the predictions obtained from empirical models \cite{tracey2013application, wang2017physics, wu2018physics, ling2016reynolds, kaandorp2020data, mcconkey2022deep}. These studies have mainly focused on obtaining ways to correct the RST prediction \cite{cruz2019use, weatheritt2016novel,weatheritt2017development,schmelzer2020discovery, amarloo2022frozen}, or to modify the available empirical models \cite{duraisamy2021perspectives, singh2017machine, holland2019field}. Taking into consideration a different approach, CFD-driven models \cite{zhao2020rans, saidi2022cfd} have shown promising results in finding reliable variants of RANS turbulence modelling. Hence, CFD-driven models can guarantee consistency and robustness in \textit{a posteriori} results (i.e. by a model-consistent training method \cite{duraisamy2021perspectives}), as opposed to other data-driven RANS models (i.e. by a \textit{a priori}-training method \cite{duraisamy2021perspectives}) since they are optimised while performing simulations to guarantee the improvement of the results obtained by new models.

% \subsection{CFD-driven literature}
In order to improve these models by a CFD-driven approach, it is necessary to solve a complex optimisation problem. In this regard, the use of optimisation algorithms that aim to minimise or maximise one or more functions in a multi-dimensional parameter space is essential. Some of the commonly used optimisation algorithms include slope followers \cite{barzilai1988two}, simplex methods \cite{nelder1965simplex}, multi-objective evolutionary \cite{coello2007evolutionary}, and particle swarm algorithms \cite{eberhart1995particle}, among others. However, these algorithms require the evaluation of an objective function, which can be computationally expensive if CFD simulations are involved, especially in cases where a large number of test configurations are required.

To address this issue, development has been made towards the use of a relatively smaller set of simulations to create computationally efficient surrogate models, also known as response surfaces, that can then be fastly optimised \cite{forrester2008engineering}. Response surfaces are mathematical models that approximate the behaviour of the objective function in the parameter space and can be used to predict the function values at untested configurations. This approach has been applied to various engineering applications, such as optimising complex internal-flow systems based on ultrasonic flow metering \cite{rincon2022turbulent, rincon2023validating}, improving the efficiency of gas cyclones \cite{singh2017multi}, optimising the aero-structural design of plane wings \cite{lam2009coupled}, and improving the performance of ground vehicles \cite{urquhart2020aerodynamic}, among others.

There are only a few studies that have investigated the use of optimisation algorithms and data-driven models for the improvement of the RANS turbulence models. Reference ~\cite{zhao2020rans} combined CFD-driven optimisation with gene expression programming to obtain a correction for the RST modelled by the $k-\omega$ SST model. They showed that the CFD-driven model had an improved performance compared to the data-driven model trained on the same case. They concluded that CFD-driven models have a great potential for developing reliable improved RANS models even though their new model is limitedly optimised for wake mixing flow. 
In another study, Ref.~\cite{saidi2022cfd} used response surfaces to find the best linear combination of candidate functions to correct the RST values modelled by the $k-\omega$ SST model for the cases with flow separation. They also showed that a CFD-driven approach yields models that perform better than the data-driven models trained on the same set of data. 
In Ref. \cite{huang2021bayesian}, the authors resort to Kriging and obtained wall models for boundary layer flows subjected to system rotation in an arbitrary direction. 
The model was shown to predict deviations in the mean flow from the equilibrium law of the wall.
In Ref. \cite{xiang2021neuroevolution}, the authors employed an evolutionary neural network and arrived at numerical strategies for the pressure Poisson equation with density discontinuities. 
Furthermore, the CFD-driven methodology is extended to a multi-objective optimisation for coupled turbulence closure models by \cite{waschkowski2022multi}. However, the application of surrogate-based optimisation (SBO) and Bayesian optimisation methods \cite{queipo2005surrogate} to improve complex mathematical tools such as RANS turbulence modelling has not been widely explored. In this regard, SBO and Bayesian optimisation methods show potential in evaluating problems based on design and analysis of computer experiments (DACE) \cite{sacks1989designs} with a low-to-medium number of parameters to optimise.

One of the most important topics in data-driven RANS modelling is the generalisability of the new models for unseen cases \cite{sandberg2022machine}. It has been suggested that using a multi-case CFD-driven approach to consider different turbulent phenomena during the optimisation process can help with the generalisability of the new model \cite{fang2023toward}. However, even these new models are still specific to either wall-free or wall-bounded flows; therefore, the generalisability problem requires further investigation.

The generalisability is the ability of data-driven models to perform well both with unseen cases (out of the training cases range) and with canonical cases like channel flow. The combination of a conventional progressive approach with data-driven approaches has been proposed to address this issue \cite{bin2022progressive}. In the progressive approach, the starting point is a baseline model that is already performing adequately for simple flows and adds more complexity step by step without violating the model's performance for flow cases that the model was calibrated against. In this study, we use the progressive approach to add the capability of secondary flow reconstruction to a linear eddy-viscosity model without violating its successful performance in a channel flow simulation.

\begin{figure}[t]
	\centering
	\includegraphics[width=1\linewidth]{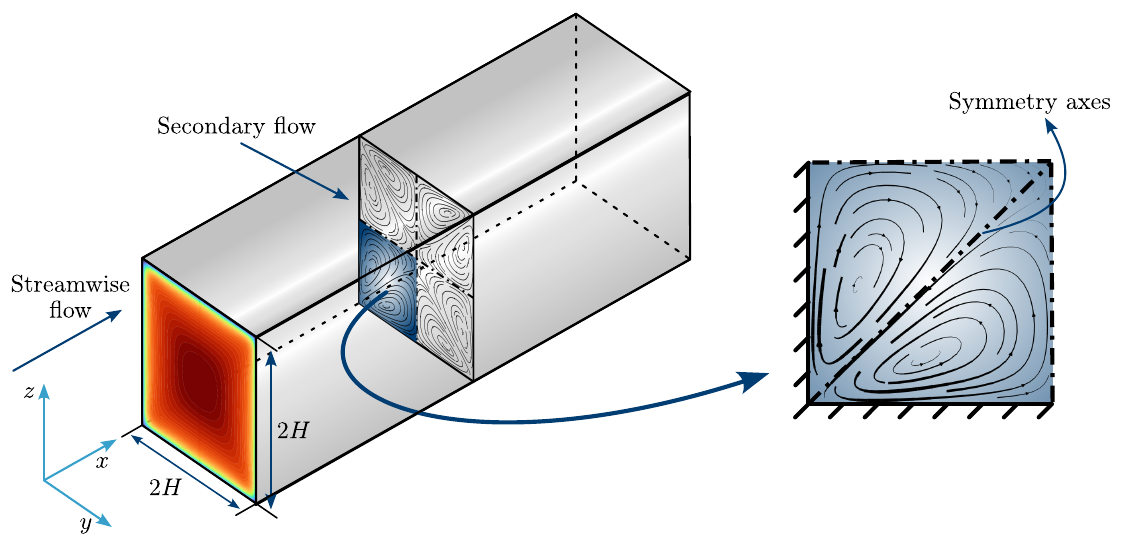}
	\caption{Flow characteristics of a duct flow with AR $=1$ and hight and width of 2H. Note the formation of secondary motions and their symmetry.}
	\label{fig:Fig1}
\end{figure}

Since the CFD-driven technique ensures the consistency of the \textit{a posteriori} results and the progressive approach aids with the generalisability of the new models, in this study we combine the CFD-driven surrogate and Bayesian optimisation technique with the progressive approach to obtain a \blue progressively data-augmented explicit algebraic Reynolds stress model (PDA-EARSM) \black to the $k-\omega$ SST \cite{menter1994two} model for exclusively predicting secondary flows. We likewise investigate the potential of a progressive approach in the development of generalisable CFD-driven RANS models. Since standard linear eddy-viscosity models have difficulty predicting secondary flows \cite{nikitin2021prandtl}, we use the Pope's decomposition of RST \cite{pope1975more} to add a non-linear term of the RST to the model, and we optimise the new models for the prediction of secondary flows induced by a duct flow with an aspect ratio (AR) of 1 (squared) at bulk Reynolds number of $3500$ (depicted in Fig. \ref{fig:Fig1}). The new models' boundary-layer prediction is tested on the channel flow to ensure that they do not affect the successful performance of the original model. Considering the generalisability, the new models are verified on duct flow cases with different Reynolds numbers and aspect ratios. Furthermore, we test the new models against an extremely disadvantageous case regarding accurate flow prediction by RANS models, we verify the models in a case where the secondary flow is induced by roughness patches in a channel flow \cite{forooghi2020roughness, amarloo2022secondary} with a nominally infinite Reynolds number. \blue Finally, we test the performance of the models against two novel data-driven EARSMs and the traditional BSL-EARSM based on the constitutive relation of Wallin and Johansson \cite{wallin2000explicit}. \black

\section{Methodology}
\label{sec:methodology}

This section presents the structure of the progressive correction model for RANS modelling and the optimisation technique for training the correction model.
By using the Reynolds decomposition of velocity and pressure, the Navier-Stokes equation for an incompressible steady flow can be written as \cite{pope2000turbulent}

\begin{equation}
\label{eq:ransDC}
u_i = \langle u_i \rangle + u^{\prime}_i, \hspace{0.2cm}  p = \langle p \rangle + p^{\prime},   
\end{equation}

\begin{equation}
\label{eq:ransEq}
\partial_i \langle u_i \rangle = 0, \hspace{0.2cm} \partial_j (\langle u_i \rangle \langle u_j \rangle) = -\frac{1}{\rho}\partial_i \langle P \rangle + \partial_j \left(\nu \partial_j \langle u_i \rangle - A_{ij}\right),   
\end{equation}
where $i, j = 1, 2, 3$ are indicating the streamwise ($x$), spanwise ($y$), and vertical ($z$) directions, respectively. $u_i$ and $p$ are velocity and pressure decomposed to a temporal mean term, indicated by $\langle \cdot \rangle$, and fluctuations, indicated by $\cdot^{\prime}$. The kinematic viscosity is denoted by $\nu$, and $\rho$ is the fluid density. $ A_{ij} =\langle u^{\prime}_{i}u^{\prime}_{j} \rangle - \frac{1}{3}\langle u^{\prime}_{k} u^{\prime}_{k} \rangle \delta_{ij}$ is the anisotropic part of RST, and pressure is modified with the isotropic part of the RST as $\langle P \rangle = \langle p \rangle + \frac{1}{3}\rho\langle u^{\prime}_{i} u^{\prime}_{i} \rangle$.

In this study, the $k-\omega$ SST \cite{menter1994two} model is used as a baseline model to be progressively corrected to predict secondary flows. In the standard $k-\omega$ SST model, $A_{ij}$ is modelled as
\begin{equation}
\label{eq:eddyViscosityModel}
A^{BL}_{ij} = -2\nu_t S_{ij},
\end{equation}
where $S_{ij} = \frac{1}{2}(\partial_i\langle u_j \rangle + \partial_j \langle u_i \rangle)$ is the strain rate tensor, $\nu_t$ is the turbulent viscosity which is calculated by values of turbulent kinetic energy (TKE) (i.e. $k$), and the specific dissipation rate (i.e. $\omega$), which are modelled by the original two-equation model \cite{menter2003ten}.

\subsection{Progressively data-augmented EARSM}
We extend the structure of the linear eddy-viscosity model (Eq.\ref{eq:eddyViscosityModel}) by following Pope's decomposition \cite{pope1975more} of the RST and considering the two first terms of the decomposition,
\begin{equation}
    A_{ij} = 2k\alpha^{(1)} \frac{S_{ij}}{\omega} + 2k\alpha^{(2)}  \frac{S_{ik}\Omega_{kj}-\Omega_{ik}S_{kj}}{\omega^2},
    \label{eq:popeDecomposition}
\end{equation}
where $\Omega_{ij} = \frac{1}{2}(\partial_i\langle u_j \rangle - \partial_j \langle u_i \rangle)$ is the rotation rate tensor, and $\alpha^{(n)}$ are unknown functions of 5 invariants defined as,

\begin{equation}
    I_{1} = \frac{\text{tr}(S_{ik}S_{kj})}{\omega^{2}},
    I_{2} = \frac{\text{tr}(\Omega_{ik}\Omega_{kj})}{\omega^{2}},
    I_{3} = \frac{\text{tr}(S_{ik}S_{km}S_{mj})}{\omega^{3}},
    I_{4} = \frac{\text{tr}(\Omega_{ik}\Omega_{km}S_{mj})}{\omega^{3}},
    I_{5} = \frac{\text{tr}(\Omega_{ik}\Omega_{km}S_{ml}S_{lj})}{\omega^{4}}.
    \label{eq:invariants}
\end{equation}

A comparison between Eq.~\ref{eq:eddyViscosityModel} and Eq.~\ref{eq:popeDecomposition} shows that the $k-\omega$ SST model is already providing the first term in Pope's decomposition of the RST (i.e., the turbulent viscosity); therefore following the progressive approach, only the second term is trained in this study. The reason behind this decision is to preserve the original performance of the $k-\omega$ SST model in the prediction of the turbulent viscosity (i.e., $\nu_t$) while the second basis tensor's coefficient is determined with the exclusive purpose of secondary flow prediction. Hence, the new RST model can be written as
\begin{equation}
    A_{ij} = -2\nu_{t}\left( S_{ij} - \alpha^{(2)} \frac{S_{ik}\Omega_{kj}-\Omega_{ik}S_{kj}}{\omega} \right),
     \label{eq:rstModel}
\end{equation}
where $\nu_t$ and $\omega$ are modelled by the standard $k-\omega$ SST model, where Eq.~\ref{eq:rstModel} is used for the production term by the Reynolds stress tensor instead of Eq.~\ref{eq:eddyViscosityModel}, and the unknown function of $\alpha^{(2)}$ is determined by the CFD-driven optimisation technique. It should be mentioned that the linear part of the new model is treated implicitly as turbulent viscosity, and the non-linear part of the RST is added explicitly to the RANS equations.
The assumption behind the progressive approach is that using only the second basis tensor does not affect the incompressible parallel shear flow, either in the momentum equation or via production in the $k$-equation; therefore, the performance of $k-\omega$ SST is preserved in cases where secondary flow is not present.

Inspired by a sparse regression of candidate functions (SpaRTA) \cite{schmelzer2020discovery} used for $\alpha^{(n)}$, we use a set of candidate functions to describe $\alpha^{(2)}$ as
\begin{equation}
    \label{eq:regressionFunction}
    \alpha^{(2)} = \theta_0 +  \sum_{i=1}^{20} \theta_i \mathcal{D}_i,
\end{equation}
\begin{equation}
    \label{eq:candideFunctions}
    \begin{split}
        \mathcal{D} = \{& I_1, I_2, I_3, I_4, I_5, I_1^2, I_2^2, I_3^2, I_4^2, I_5^2, \\
                        & I_1I_2, I_1I_3, I_1I_4, I_1I_5, I_2I_3, I_2I_4,\\
                        & I_2I_5, I_3I_4, I_3I_5, I_4I_5\},
    \end{split}
\end{equation}
where $\theta_{i}$ are constant coefficients to be determined by the CFD-driven optimisation process. To achieve a more efficient sparse optimisation, the normalised candidate functions are normalised and defined as
\begin{equation}
    \label{eq:normalization}
    \mathcal{B}_i=\frac{\mathcal{D}_i - \mu_i }{\sigma_i},
\end{equation}
where $\mu_i$, $\sigma_i$ are the mean and the standard deviation of each candidate function $\mathcal{D}_i$, respectively. These statistics are calculated based on high-fidelity data from the optimisation case. Therefore, Eq.~\ref{eq:regressionFunction} is rewritten as,
\begin{equation}
    \label{eq:modelI}
    \alpha^{(2)} = C_0 +  \sum_{i=1}^{20} C_i \mathcal{B}_i,
\end{equation}
where an optimisation technique determines the coefficients $C_i$ based on the performance of the correction model for the reconstruction of the high-fidelity velocity field of an optimisation case. 

In this study, only 2D canonical flow cases are considered, since they are computationally cost-effective. However, reducing the dimensionality of the optimisation problem and using computational parallelisation is key if a complex 3D case is used in the training process.

Since considering 21 optimisation variables is impractical, making the model unnecessarily complex and exposed to solution instabilities, two approaches are chosen:
\begin{enumerate}
    \item Selecting only the first $m$ leading candidate functions.
    \item Reducing the dimensionality of the problem using a statistical technique.
\end{enumerate}
% only selecting the first $m$ leading candidate functions, and using a statistical technique to reduce the dimensionality of the problem. 

 Both of these approaches are compared, and for the purpose of dimensionality reduction, principal component analysis (PCA) is applied on the $\mathcal{B}_{i}$ functions to obtain the first $m$ principal components as,
\begin{equation}
    \varphi_{j} = \sum_{i=1}^{20} a^{(j)}_{i}\mathcal{B}_{i},
\end{equation}
where the coefficients $a^{(j)}_{i}$ are obtained by performing PCA on the high-fidelity data from the optimisation case. It should be mentioned that PCA also determines which features among all the 20 features have a higher importance in the variability of $\alpha^{(2)}$ by determining the $a_i^{(j)}$ coefficients for each of them. By considering this transformation equation, Eq.~\ref{eq:modelI} is rewritten as,
\begin{equation}
    \label{eq:modelII}
    \alpha^{(2)} = C_0 +  \sum_{i=1}^{m} C_i \varphi_i.
\end{equation}
Based on the high-fidelity data of the optimisation case, in Sec. \ref{sec:trainingProcess} it is shown that the first two principal components are enough to represent a high percentage of the variability of the dataset (i.e. $m=2$). Therefore, two different structures for $\alpha^{(2)}$ are considered:
\begin{subequations}
\label{eq:models3}
    \begin{equation}
        \alpha^{(2)}_{I} = C_0 + C_1 \mathcal{B}_1 + C_2 \mathcal{B}_2,
    \label{eq:modelI3}
    \end{equation}    
    \begin{equation}
        \alpha^{(2)}_{II} = C_0 +  C_1 \varphi_1 + C_2 \varphi_2,
    \label{eq:modelII3}
    \end{equation}
\end{subequations}
which are described as \textit{model \textbf{I}} and \textit{model \textbf{II}}, respectively. Where a unique set of three coefficients $C_0$, $C_1$, and $C_2$ are determined by a multi-objective optimisation technique for each of the models.

\subsection{Optimisation methods}

\begin{figure}[t]
	\centering
	\includegraphics[width=1\linewidth]{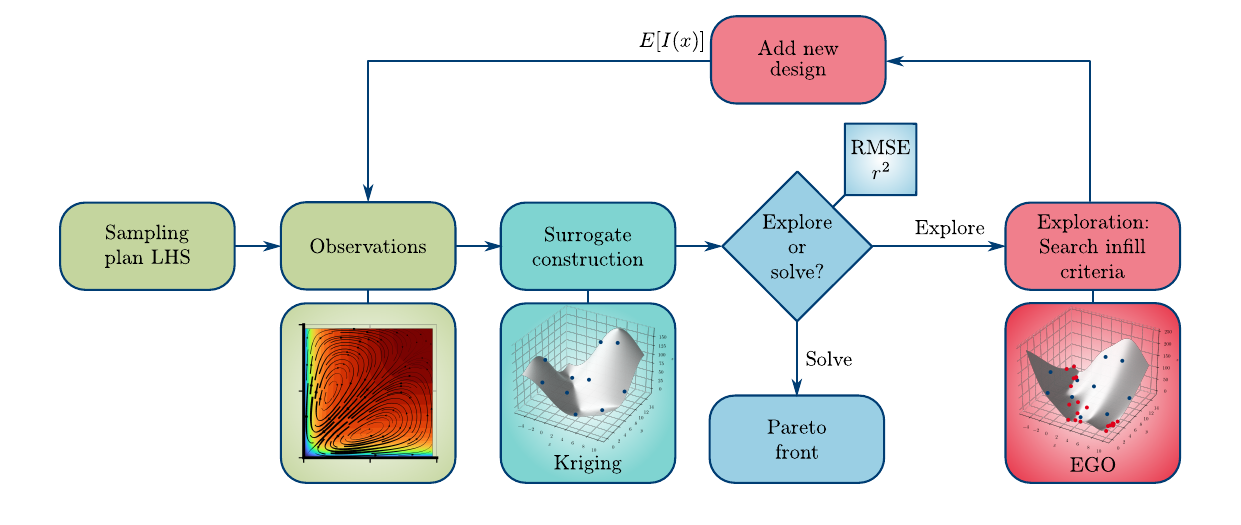}
	\caption{Optimisation strategy employed involving an initial sampling plan based on Latin hypercube sampling (LHS) which is solved by CFD. Subsequently, an initial surrogate is constructed using Kriging. Since it is likely that the initial sampling may not evaluate the surrogate in extrema, Bayesian strategies based on efficient global optimisation (EGO) and the evaluation of the expected improvement ($E\left[ I(x) \right]$) function are then applied to further explore the surrogate and improve its quality. Finally, once the quality requirements have been met, a genetic algorithm is employed to search for non-dominated optima on the surrogate model.}
	\label{fig:Fig2}
\end{figure}

To approach the optimisation problem, a surrogate based on Kriging and Gaussian processes (GPs) is built. The surrogate is based on DACE, using observations obtained from a sequence of CFD solutions and post-processing of results. SBO has shown its advantages for these types of problems, minimising the number of required observations, and allowing the use of goal-seeking algorithms such as Bayesian optimisation \cite{jones2001taxonomy}. To assess the model's performance, two main objective functions are established that involve streamwise velocity and streamwise vorticity, defining a multi-objective optimisation problem. An effective sampling plan is also outlined, and an infill criterion constrained by quality metrics is specified. Finally, the methodology is implemented using the general-purpose software OpenFOAM \cite{weller1998tensorial}. A flow diagram of the complete optimisation methodology is depicted in Fig. \ref{fig:Fig2}.

\subsubsection{Objective functions}
The principal issue that is assessed in this study is the complete lack of prediction of secondary flows in RANS turbulence models based on the Boussinesq assumption. In the chosen optimisation case, the addition of these secondary flow motions is strong enough to also incur changes in the streamwise direction of the flow. In addition, once the corrections are made, the mean streamwise flow is simultaneously modified. Since the aim of this study is to improve the overall flow prediction, it is important to not disregard possible predictions that could improve secondary flow but worsen streamwise velocity. Hence, two objective functions that physically describe the streamwise flow and the secondary flow prediction are defined and evaluated.

In this regard and to holistically evaluate the flow prediction, the normalised error of both the streamwise velocity and vorticity with respect to high-fidelity data, are evaluated and taken as objectives to be minimised. To yield a fair and accurate single value representing these quantities, the volumetric average of each field is computed as the objective functions, following
\begin{subequations}
\label{eq:objectives}
    \begin{equation}
        j_{1} = \frac{\int_{V} \left( \langle u_{1} \rangle_{\text{PDA-EARSM}} - \langle u_{1} \rangle_{\text{HF}} \right) \text{d}V}{ \int_{V} \left( | \langle u_{1} \rangle_{k-\omega} - \langle u_{1} \rangle_{\text{HF}} | \right) \text{d}V},
        \label{eq:streamwiseObjective}
    \end{equation}
    \begin{equation}
        j_{2} = \frac{\int_{V} \left( \langle \omega_{1} \rangle_{\text{PDA-EARSM}} - \langle \omega_{1} \rangle_{\text{HF}} \right) \text{d}V}{\int_{V} \left( | \langle \omega_{1} \rangle_{k-\omega} - \langle \omega_{1} \rangle_{\text{HF}} | \right) \text{d}V},
        \label{eq:vorticityObjective}
    \end{equation}
\end{subequations}
where $| \cdot |$ is the absolute value operation, $\langle \omega_i \rangle = \varepsilon_{ilk}\partial_l \langle u_k \rangle$ is the vorticity of the mean flow, where $\varepsilon_{ilk}$ is the alternating unit tensor, HF stands for high fidelity, and $k-\omega$ refers to the solution of the standard $k-\omega$ SST model. The definition of these functions yields a value of 1 when their output matches the prediction of $k-\omega$ SST and 0 when the predictions match the high-fidelity data. Subsequently, to evaluate the overall results of the method, a global fitness function is defined as
\begin{equation}
    J = \frac{1}{2}\left( j_{1} + j_{2} \right).
\end{equation}
\subsubsection{Sampling plan}
Two main algorithms are used in order to sample observations and data in this study: Monte Carlo and Latin hypercube. Monte Carlo sampling is used when extracting a test subset to compute the surrogate quality. Since the chosen sampling plan for the initial observations is oriented to be \textit{space-filling}, Latin Hypercube Sampling (LHS) \cite{mckay2000comparison} with optimised design by the enhanced stochastic evolutionary (ESE) algorithm  \cite{jin2005efficient, damblin2013numerical} is used to obtain an initial sample of the surrogate with a minimum number of observations, and simultaneously, representing the real variability of the parameters. This deterministic sampling technique is a type of stratified Monte Carlo that divides each dimension space representing a variable into $n_{0}$ sections, and only one point is placed in each partition. The initial sample by LHS with ESE follows $n_{0}=30\mathcal{K}$, where $n_{0}$ is the initial number of observations, and $\mathcal{K}$ is the number of design variables.

\subsubsection{Surrogate construction}
In the context of a sparse optimisation problem, surrogates are commonly used to approximate a function $y=f(x)$ based on known observations. In this study, the Kriging method is chosen for generating the surrogate based on CFD-driven observations. Kriging is one of the most common surrogate construction methods due to its well-known implementation, ability to compute uncertainties and speed. Furthermore, it has been proven useful in physical, uncertainty quantification, and engineering applications \cite{rincon2023validating, kawai2014kriging}. Kriging interpolates the observations as a linear combination of a deterministic term and a stochastic process, which is represented by
\begin{equation}
    \hat{f}(x) = \sum_{i=1}^{k} \beta_{i} f_{i}(x) + Z(x),
\end{equation}
where $\hat{f}(x)$ is the surrogate prediction, $\beta$ is a linear deterministic model, $f(x)$ is the known function, and $Z(x)$ is the realisation of a stochastic process with zero mean and spatial covariance function given by
\begin{equation}
    \text{cov}\left[ Z \left( x^{(i)} \right), Z \left( x^{(j)} \right) \right] = \sigma^2 R \left( x^{(i)}, x^{(j)} \right).
\end{equation}
Here, the spatial correlation function $R$ determines how smooth the Kriging model is, how easily the response surface can be differentiated, and how much influence the nearby sampled points have on the model. In this study, the spatial correlation is defined following the squared exponential (Gaussian) function as
\begin{equation}
    \prod_{i=1}^{nx}\exp{\left[ -\theta_{l} \left( x_{l}^{(i)} - x_{l}^{(j)} \right)^{2} \right]}, \quad \forall \theta_{l} \in \mathbb{R}^{+},
\end{equation}
where the correlation scalar $\theta_{l}$ is used to define the variance of a Gaussian process at each point, with higher values indicating a stronger correlation between points. By maximising the maximum likelihood estimation, optimal values for hyper-parameters as $\theta_{l}$, mean, and standard deviation can be found \cite{SMT2019}.

\subsubsection{Quality metrics}
The quality of the surrogate refers to how accurately it approximates the true function being modelled. In order to evaluate this quality, a number of random test observations $n_{t} = n_{0} \ x \xrightarrow{} y$, are taken. where $n_{t}$ is a function of the number of initial observations. Once the test observations have been generated, they are used to compare the surrogate's predictions to the true values of the function being modelled. The root-mean-squared error (RMSE) and Pearson's correlation coefficient squared ($r^2$), defined as
\begin{subequations}
    \begin{equation}
        \text{RMSE} = \sqrt{\frac{\sum_{i=1}^{n_{t}}\left( y^{(i)} - \hat{y}^{(i)} \right)^{2}}{n_{t}}},
    \end{equation}    
    \begin{equation}
        r^{2} = \left( \frac{\text{cov}(y, \hat{y})}{\sqrt{\text{var}(y)\text{var}(\hat{y})}} \right)^{2},
    \end{equation}
\end{subequations}
are two common metrics used to evaluate the quality of surrogates. While the RMSE measures the difference between the surrogate's predictions and the true values, $r^2$ measures the strength of its linear relationship. A high RMSE and a low $r^2$ indicate that the surrogate is performing poorly and needs to be refined, while a low RMSE and a high $r^2$ indicate that the surrogate is performing accurately and can be used with confidence. Hence, these values serve as reference points to determine whether the surrogate model is accurate enough to be used in the optimisation process.

In this study, thresholds of 0.2 for RMSE and 0.8 for $r^2$ are established as indicators of good surrogate quality, based on previous studies \cite{sobester2008engineering, hastie2009elements}. 

\subsubsection{Infill space exploration}
Efficient global optimisation (EGO) based on Bayesian optimisation strategies is used to improve the accuracy of surrogate models and decrease overall uncertainty \cite{jones2001taxonomy, jones1998efficient}. This is achieved by exploring the surrogate beyond the initial sampling. EGO is a well-known algorithm that employs both local and global searches to find the optimal solution by means of the expected improvement ($E[I(x)]$) function as a key metric to direct its search. The function calculates the potential improvement that can be obtained by evaluating a new observation point based on the current best solution and the overall uncertainty of the surrogate model \cite{mockus1978application}. The implementation of this function provides the necessary sparsity-promoting behaviour to explore the design space in regions where their initial sampling was not sufficient. The expected improvement function is defined as
\begin{equation}
    E[I(x)] = \left( f_{\text{min}} - \mu(x) \right) \Phi \left( \frac{f_{\text{min}} - \mu(x)}{\sigma(x)} \right) + \sigma(x) \phi \left( \frac{f_{\text{min}} - \mu(x)}{\sigma(x)} \right),
\end{equation}
where $f_{\text{min}} = \min{Y}$, and $\Phi(\cdot)$ and $\phi(\cdot)$ are respectively the cumulative and probability density functions of $\mathcal{N}(0,1)$, following the distribution $\mathcal{N}(\mu(x),\sigma^{2}(x))$. Using EGO to explore the surrogate beyond the initial sampling aids in decreasing the overall uncertainty and improving the precision of the surrogate model, especially in the unexplored and extreme regions of the design space. Consequently, the algorithm can effectively balance local and global searches and locate the optimal solution more proficiently. Hence, the following sampling point is determined by
\begin{equation}
    x_{n+1} = \arg \max_{x} \left( E[I(x)] \right).
\end{equation}
In order to fully explore the design space, new data is collected for each objective function separately. This involves taking one sample of $n_{\text{EGO}}=n_{0} \ x \xrightarrow{} y$ for each objective, resulting in a total of $180$ new sampled points ($n_{0}$ for each of the objectives). With the addition of the initial LHS, the total number of observations sums to 270 per model, yielding a cost-efficient computational problem to be analysed by Kriging. However, since the design space explored is large and the use of Bayesian optimisation might sample observations in very close proximity, the gradients between points may be high, yielding possible noise. Therefore, the surrogate is regularised following the methodology by \cite{bouhlel2019python}, where noise is considered to be Gaussian distributed and its homoscedastic variance is evaluated based on the performed observations to obtain a smooth response surface.

% the methodology evaluates this noise and applies L1 regularisation to obtain a smooth response surface.

\subsubsection{Multi-objective solution}
The Kriging method offers a significant benefit since it allows the application of algorithms that can thoroughly explore the objective space. Consequently, the objective of this study is to find the optimal solution for the surrogate model. To achieve this, the multi-objective evolutionary algorithm (MOEA) NSGA-II has been chosen. This algorithm is known for its versatility, fast and efficient convergence, and ability to handle non-penalty constraints. It also has a wide-range search for solutions, which implies that it can explore the objective space more thoroughly and is less likely to get grounded in local optima \cite{deb2002fast}. By using the MOEA NSGA-II algorithm, the improved solutions are ensured to be non-dominated, meaning that they are not worse than any other feasible solution in all objective functions. This also ensures that the solutions avoid local minima and are highly likely to discover the global optimum for the sampled response surface. This makes the algorithm able to search a large portion of the objective space and find the best possible solution for the surrogate model while avoiding suboptimal solutions.

\subsection{High-fidelity data}
Since the main purpose of this study is to obtain a PDA-EARSM for capturing the secondary flow, the DNS data of a canonical case with this flow characteristic is chosen. A duct flow case of AR $=1$ and bulk Reynolds number of $\text{Re}_{b}=3500$ is used for the training process (depicted in Fig.~\ref{fig:Fig1}). The DNS data is obtained from Ref.~\cite{pinelli2010reynolds} curated by Ref.~\cite{mcconkey2021curated}. 
Following the progressive approach, the trained PDA-EARSMs are tested on two cases of channel flow with friction Reynolds number of $\text{Re}_\tau=395$ and $\text{Re}_\tau=5200$ for which data is obtained from Refs.~\cite{moser1999direct} and \cite{lee2015direct}, respectively.
Considering the generalisability of the new models, we likewise test them on \blue four \black unseen cases containing secondary flows, including duct secondary flow cases with AR $=1$ and higher bulk Reynolds numbers of $\text{Re}_{b}=5700$ \cite{vinuesa2018secondary} \blue and $\text{Re}_{b}=10320$ \cite{huser1993direct} \black, a duct secondary flow case with AR $=3$, a lower bulk Reynolds number of $\text{Re}_{b}=2600$ \cite{vinuesa2018secondary}, and a roughness-induced secondary flow with nominally infinite Reynolds number (more information about the geometry and properties of this case is available at Refs. \cite{amarloo2022secondary,amarloo2022frozen}. 

% The CFD test case chosen to apply the methodology and build the surrogate is a duct with an aspect ratio (AR) of 1 and Re$=5600$ from \cite{pinelli2010reynolds}. The duct dimensions are such that the height is equal to the width (square cross-section), where the length is assumed to be infinite. The fluid is assumed to be incompressible and Newtonian, with a constant density and viscosity. The flow in the test case is driven by a constant pressure gradient between the inlet and outlet and cyclic boundary conditions. Due to the use of RANS, the case is simplified to speed-up simulation time by imposing symmetry boundary conditions in the spanwise and vertical directions representing $1/4^{th}$ of the total duct and creating a 1-cell-long computational domain as shown in Fig. \ref{fig:Fig3}.

\section{Results and discussion}
\label{sec:results}
In this section, the results are presented in two subsections. In the first one, the results of the optimisation process and training of the two best PDA-EARSMs are presented, whereas, in the second, the performance of the trained models is evaluated on the test cases with different geometries, Reynolds numbers, and boundary conditions. Associated contours and velocity profiles accompany all results which are similarly described and discussed, followed by their corresponding figures.

\subsection{Surrogate-based optimisation}
\label{sec:trainingProcess}
\subsubsection{Optimisation case: $k-\omega$ SST performance}
\begin{figure}[H]
	\centering
	\includegraphics{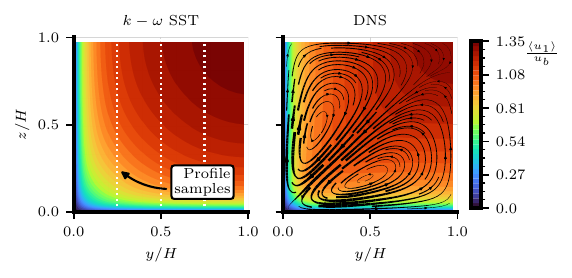}
	\caption{Contours of streamwise velocity and secondary flow streamlines for duct flow optimisation case at AR $=1$ and $\text{Re}_{b}=3500$ for RANS $k-\omega$ SST (left) and DNS (right) data. DNS data obtained from Ref.~\cite{pinelli2010reynolds}. The dotted white lines denote the location of the velocity profiles evaluated throughout this study to perform a quantitative comparison. It should be noted the varying thickness of the secondary flow streamlines denotes their magnitude.}
	\label{fig:Fig3}
\end{figure}

It is essential to qualitatively indicate the original performance of $k-\omega$ SST against high-fidelity data prior to any modifications. For both fidelity levels, the results for the square duct case show the progressive deceleration of the streamwise flow as it approaches the walls of the duct, exhibiting higher gradients in the near-wall regions as a consequence of the boundary layer and the fully developed turbulent flow. Whereas this behaviour can be predicted by $k-\omega$ SST, the presence of a secondary flow predicted by DNS is completely neglected by the RANS model (Fig. \ref{fig:Fig3}), as expected \cite{nikitin2021prandtl}. Two antisymmetric rotational regions are generated diagonally at the duct's vertex, preserving their symmetry at the 4 quadrants of the duct. This secondary flow is strong enough to drive the streamwise flow towards the duct vertices, skewing the streamwise velocity field in the domain through an impinging-like flow behaviour. Since $k-\omega$ SST does not predict these rotational motions, the streamwise velocity distribution is not skewed and the spanwise and vertical components of the velocity are zero. 

% THIS IS WEIRD 
% In addition, a qualitative analysis is likewise performed evaluating the velocity profiles at $y/H \in [0.25, 0.5, 0.75]$ and comparing them for all tested cases as depicted in Fig. \ref{fig:Fig3}.

\begin{figure}[H]
	\centering
    \includegraphics{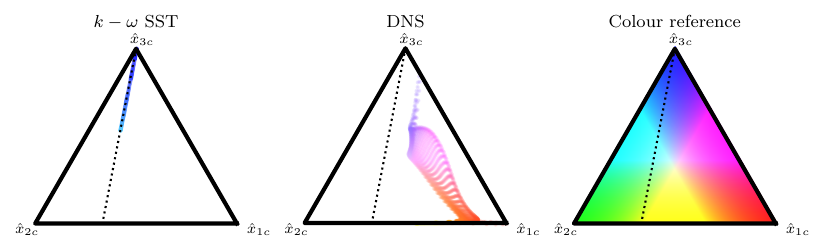}
	\caption{Barycentric representation of RSTs' shape for $k-\omega$ SST (left) and DNS (centre) for training duct flow case with AR $=1$ at $\text{Re}_{b}=3500$. DNS data obtained from Ref.~\cite{pinelli2010reynolds}. The plain-strain limit is represented by the dotted line, and the colour representation (right) follows the work by \cite{emory2014visualizing}.}
	\label{fig:Fig4}
\end{figure}

Figure~\ref{fig:Fig4} compares the barycentric plots representing the shape of RSTs for the optimisation case. The location of each point is calculated based on the eigenvalues of the normalised anisotropic part of the RST (more information about barycentric plots available at Refs.~\cite{hornshoj2021quantifying, emory2014visualizing}). While $k-\omega$ SST yields all results within the plain-strain limit as expected from a linear eddy-viscosity RANS model, DNS shows a more complex stress anisotropy distribution as a combination of states toward one-component turbulence fluctuations $\hat{x}_{1_{c}}$ (also known as rod-like or cigar-like turbulence) and an offset from the plain-strain limit with a tendency towards isotropic (or spherical) turbulence $\hat{x}_{3_{c}}$.

\subsubsection{Optimisation case: PDA-EARSMs} 
Since $k-\omega$ SST is not able to predict the secondary flow of the optimisation case, CFD-driven optimisation is applied following the models described in Eqs.~\ref{eq:modelI3} and \ref{eq:modelII3} focusing on predicting secondary flows and correct streamwise velocity.

\begin{figure}[H]
     \centering
     \begin{subfigure}[t]{0.495\linewidth}
        \centering
        \includegraphics[width=1\textwidth]{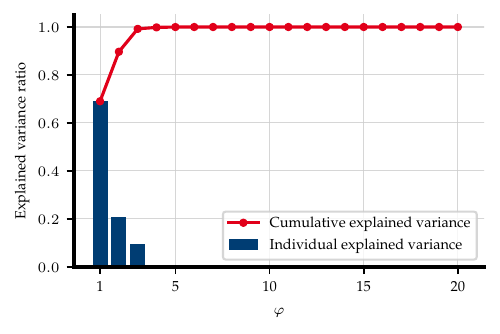}
        \caption{Explained variance ratio of each principal component.}
        \label{fig:Fig5a}
     \end{subfigure}
     \hfill
     \begin{subfigure}[t]{0.495\linewidth}
        \centering
        \includegraphics[width=1\textwidth]{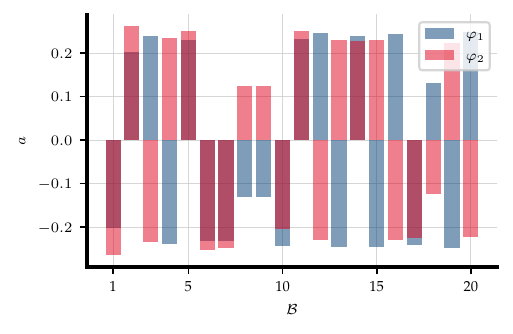}
        \caption{Coefficients associated with the first two principal components.}
        \label{fig:Fig5b}
     \end{subfigure}
     \caption{PCA results of the candidate functions calculated by DNS data of the optimisation case.}
     \label{fig:Fig5}
\end{figure}

Regarding model \textbf{II}, PCA is applied to the set of candidate functions (listed in Eq.~\ref{eq:candideFunctions}) using the high-fidelity data of the optimisation case. Figure~\ref{fig:Fig5} presents the PCA results and shows that only the first 2 principal components ($\varphi_{1}$ and $\varphi_{2}$) yield an explained variance ratio of 0.90 (Fig.~\ref{fig:Fig5a}), where $\varphi_{1}$ is the main principal component explaining the variability. Figure~\ref{fig:Fig5b} presents the coefficients corresponding to the two first principal components used for model \textbf{II}.

% Hence, the two first principal components are selected for model \textbf{II} (described in Eq.~\ref{eq:modelII}). \textcolor{blue}{ For a fair comparison between model \textbf{I} and model \textbf{II}, only the first two candidate functions are chosen for model \textbf{I} (described in Eq.~\ref{eq:modelI})}; therefore, both models contain 3 coefficients as ?
% \begin{subequations}
% \label{eq:models3}
%     \begin{equation}
%         \alpha^{(2)}_{I} =C_0 + C_1 \mathcal{B}_1 + C_2 \mathcal{B}_2,
%     \label{eq:modelI3}
%     \end{equation}    
%     \begin{equation}
%         \alpha^{(2)}_{II} = C_0 +  C_1 \varphi_1 + C_2 \varphi_2,
%     \label{eq:modelII3}
%     \end{equation}
% \end{subequations}
% where $C_0$, $C_1$, and $C_2$ are the optimisation (or design) variables.
As mentioned in Sec.~\ref{sec:methodology}, the two first principal components for model \textbf{II} are chosen. To compare model \textbf{I} in an analogous manner, the two first leading candidate functions are likewise chosen where the coefficients of these two models are determined by surrogate and Bayesian optimisation.

The surrogate model is generated using the RANS simulations results to find the optimal values of the optimisation variables. Regarding the optimisation process and since cases with high values of the objective functions are not of interest, the MOEA optimisation is performed by imposing the constraints $j_{1} \leq 0.5$ and $j_{2} \leq 0.4$, therefore, the analysed non-dominated solutions only take place in the regions of highest interest and potential global minimisation. To yield a smooth non-dominated solution front, 5000 samples are required by the MOEA in this study. This showcases the cost advantages of surrogates, where 270 CFD observations are required, yielding a cost reduction of 18.4 times.

Regarding both models' results, an optimal design space of $[-2, -1, -1] \leq C_{i} \leq [0, 1, 1]$ is chosen for model \textbf{I}, and an optimal design space of $[-1.75, -0.25, -0.25] \leq C_{i} \leq [-1.25, 0.25, 0.25]$ is chosen for model \textbf{II} after various progressive surrogate iterations. The iterations re-define the design space limits by analysing the non-dominated values of the coefficients after optimisation. If these values reach the imposed limits, the objective space is reset, and the design space is adapted.

% Contour plots in Fig.~\ref{fig:Isurrogate} illustrate how the optimisation variables affect the objective functions.  and quality results of the surrogate in relation to optimisation results. The Pareto front solution of both objective functions is displayed with its uncertainty and numerical validation. In addition, a qualitative and quantitative analysis of the prediction of secondary flow is conducted for the two best models found, simultaneously comparing the errors of streamwise velocity and vorticity. Finally, the numerical verification of the two best models is presented for various test cases with different geometries, Reynolds numbers, and boundary conditions. Associated contours and velocity profiles accompany all results and are similarly described and discussed, followed by their corresponding figures.

\begin{figure}[H]
     \begin{subfigure}[t]{1\linewidth}
        \centering
        \includegraphics[width=0.5\textwidth]{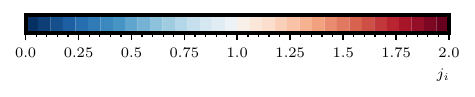}
     \end{subfigure}
     \\
     \centering
     \begin{subfigure}[t]{0.495\linewidth}
        \centering
        \includegraphics[width=1\linewidth]{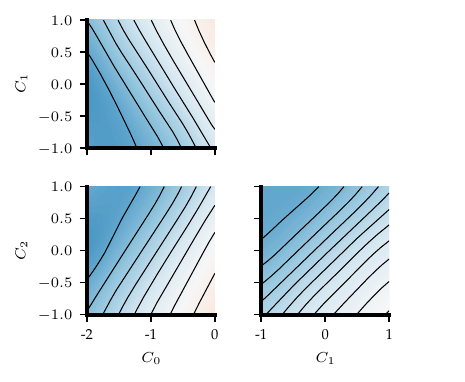}
        \caption{Surrogate for $j_{1}$.}
        \label{fig:Fig6a}
     \end{subfigure}
     \hfill
     \begin{subfigure}[t]{0.495\linewidth}
        \centering
        \includegraphics[width=1\linewidth]{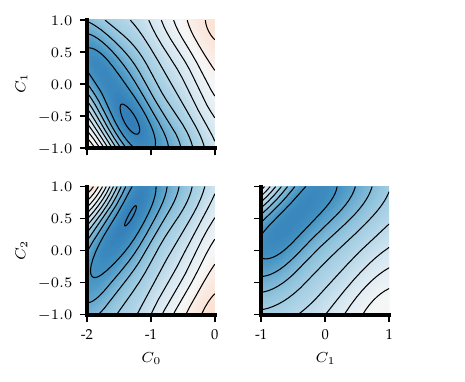}
        \caption{Surrogate for $j_{2}$.}
        \label{fig:Fig6b}
     \end{subfigure}
     \caption{Surrogate contour visualisation for model \textbf{I}. It should be noted that each pair of variables is shown by holding the other variable at the mid-value of their range.}
     \label{fig:Fig6}
\end{figure}

\paragraph{Model \textbf{I}}
Contour plots in Fig.~\ref{fig:Fig6b} illustrate how the optimisation variables affect the objective functions. On the one hand, the surrogate visualisation for $j_{1}$ (Fig.~\ref{fig:Fig6a}) shows a quasi-linear tendency for all variables. The global minimum for $j_{1}$ is shown towards negative values for all $C_{i}$ without clear visualisation of extrema. On the other hand, there is a clear global minimum seen for $j_{2}$ values, shown by a non-linear surrogate for $C_{0}$ and $C_{1}$, where quasi-linear behaviour is predicted by the relationship between $C_{1}$ and $C_{2}$ (Fig.~\ref{fig:Fig6b}). It is important to highlight that, although the lowermost values for the coefficients predict an improvement for $j_{1}$, the tendency is not completely followed by $j_{2}$ prediction. This implies that a more accurate prediction of the streamwise flow does not necessarily correlate with the secondary flow prediction in this case.

\begin{figure}[H]
     \centering
     \begin{subfigure}[t]{0.495\linewidth}
        \centering
        \includegraphics[width=1\textwidth]{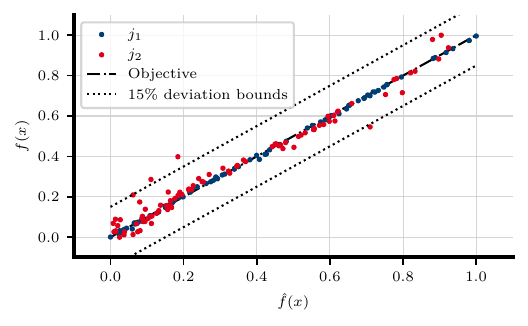}
        \caption{Surrogate quality.}
        \label{fig:Fig7a}
     \end{subfigure}
     \hfill
     \begin{subfigure}[t]{0.495\linewidth}
        \centering
        \includegraphics[width=1\textwidth]{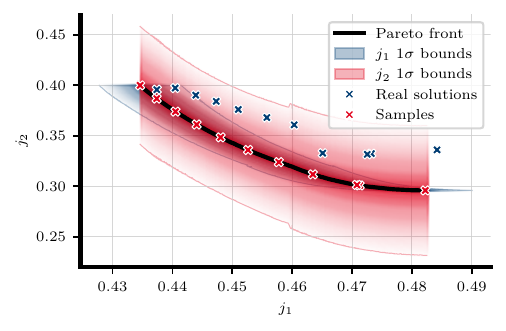}
        \caption{Pareto front.}
        \label{fig:Fig7b}
     \end{subfigure}
     \caption{Surrogate quality against test set (Fig. \ref{fig:Fig7a}), and Pareto-front validation with $1\sigma$ uncertainty bounds (Fig. \ref{fig:Fig7b}) for model \textbf{I}.}
     \label{fig:Fig7}
\end{figure}

Regarding the surrogate quality (Fig. \ref{fig:Fig7a}), and Pareto front prediction and numerical validation (Fig.~\ref{fig:Fig7b}), the surrogate predicts the test set with great accuracy for the whole design space tested, which is likewise reflected by a highly accurate Pareto front prediction, where all numerical tests simulated are able to predict the objective space within $1\sigma$ uncertainty bounds for both objective functions.

\begin{figure}[H]
     \begin{subfigure}[t]{1\linewidth}
        \centering
        \includegraphics[width=0.5\linewidth]{Fig6_Colourbar.pdf}
     \end{subfigure}
     \\
     \centering
     \begin{subfigure}[t]{0.495\linewidth}
        \centering
        \includegraphics[width=\textwidth]{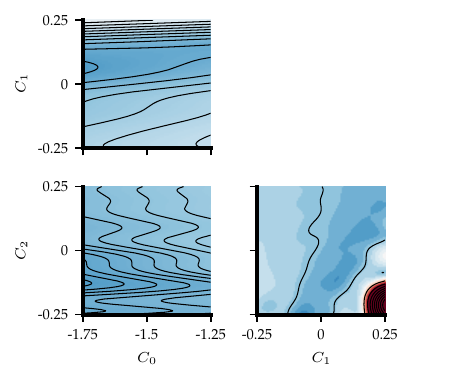}
        \caption{Surrogate for $j_{1}$.}
        \label{fig:Fig8a}
     \end{subfigure}
     \hfill
     \begin{subfigure}[t]{0.495\linewidth}
        \centering
        \includegraphics[width=\textwidth]{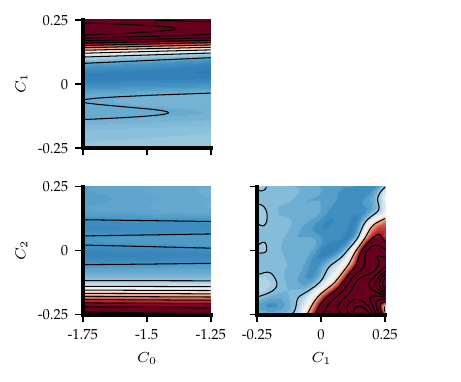}
        \caption{Surrogate for $j_{2}$.}
        \label{fig:Fig8b}
     \end{subfigure}
     \caption{Surrogate contour visualisation for model \textbf{II}. It should be noted that each pair of variables is shown by holding the other variable at the mid-value of their range.}
     \label{fig:Fig8}
\end{figure}

\paragraph{Model \textbf{II}}
In contrast with model \textbf{I}, the surrogate visualisation for both objective functions (Fig. \ref{fig:Fig8}) in model \textbf{II} shows a non-linear prediction for all variables. This behaviour is somewhat expected due to the added complexity of the model with the PCA. The global minimum for $j_{1}$ does not have a straightforward location. Instead, there are diverse regions where the global extrema can be found. Specifically, negative values of $C_{0}$, and near-zero values of $C_{1}$ and $C_{2}$ predict generally lower values of $j_{1}$ and $j_{2}$ (Figs. \ref{fig:Fig8a} and \ref{fig:Fig8b}). 
% Regarding $j_{2}$, there is a similar predicted behaviour compared to $j_{1}$: non-linear behaviour is shown for $C_{i}$, and similar values of the coefficients tend to predict the location of the minima (Fig. \ref{fig:Fig8b}).
Similarly to model \textbf{I}, although the objective spaces for both functions follow similar tendencies, there are visible differences in the predicted locations of the minima, reflecting once more that a more accurate prediction of the streamwise flow does not necessarily correlate with the secondary flow prediction.

\begin{figure}[H]
     \centering
     \begin{subfigure}[t]{0.495\linewidth}
        \centering
        \includegraphics[width=1\textwidth]{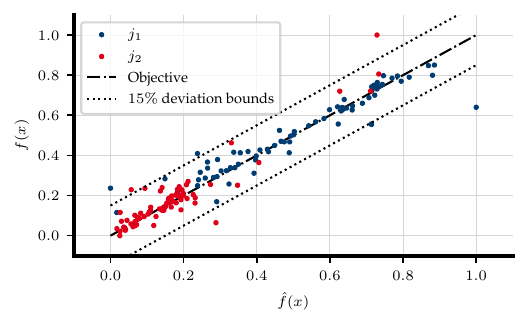}
        \caption{Surrogate quality.}
        \label{fig:Fig9a}
     \end{subfigure}
     \hfill
     \begin{subfigure}[t]{0.495\linewidth}
        \centering
        \includegraphics[width=1\textwidth]{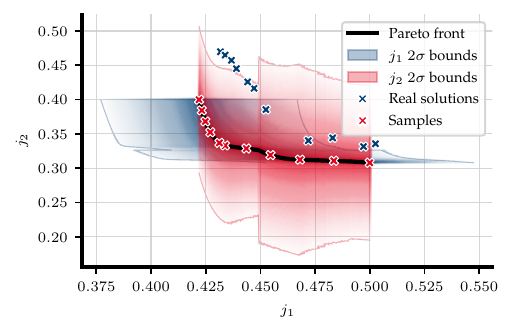}
        \caption{Pareto front.}
        \label{fig:Fig9b}
     \end{subfigure}
     \caption{Surrogate quality against test set (Fig. \ref{fig:Fig9a}), and Pareto front validation with $2\sigma$ uncertainty bounds (Fig. \ref{fig:Fig9b}) for model \textbf{II}.}
     \label{fig:Fig9}
\end{figure}

Regarding the surrogate quality (Fig. \ref{fig:Fig9a}), and Pareto front prediction and numerical validation (\ref{fig:Fig8b}), the surrogate predicts the test set with high accuracy for the whole design space tested.
% , however, some outliers can be seen in the mid-range values of the objectives. 
These lower-confidence regions are not reflected in the non-dominated solutions nor in their numerical validation. Since the expected improvement function prioritises good confidence in the near-extrema region and the complexity of the model is higher, a certain level of uncertainty is expected in far-away regions of the global minimum. Nevertheless, the surrogate shows a good level of confidence at the Pareto front (within $2\sigma$ in Fig. \ref{fig:Fig9b}).

\subsection{Comparison of models}
The final results for each model are shown in Table \ref{tab:Table1}, where the best cases per model are shown, followed by their $C_{i}$ values and surrogate quality parameters. Both models significantly improved the standard $k-\omega$ SST by $61.3\%$ and $60.2\%$ for model \textbf{I} and \textbf{II} respectively. Differences in performance for the objective values are considered negligible.
% , where model \textbf{II} is slightly showing better performance with a net $J$ gain of $1.4\%$. 
% Quality results of RMSE and $r^{2}$ are reflected as previously seen in Fig. \ref{fig:Fig7a}, and \ref{fig:Fig9a}, where model \textbf{II} displays the best overall performance. 
As seen in Fig \ref{fig:Fig7a} and \ref{fig:Fig9a}, the surrogate quality is slightly worse for model \textbf{II}. This is due to the higher complexity added by PCA and the inclusion of higher order terms in $\alpha_{2}$, yielding more non-linear behaviour in the design space and, thus, a higher RMSE and lower $r^{2}$ for model \textbf{II}.
\begin{table}[H]
\centering
\caption{Objective function results, coefficients, and quality for best models and surrogates of each approach.}
\label{tab:Table1}
\begin{tabular}{@{}lllllllllll@{}}
\toprule
Model                        & $J$   & $j_{1}$ & $j_{2}$ & $C_{0}$ & $C_{1}$ & $C_{2}$ & RMSE$_{1}$ & RMSE$_{2}$ & $r^{2}_{1}$ & $r^{2}_{2}$ \\ \midrule
\textbf{I}  & 0.387 & 0.455   & 0.319   & -1.653        & 0.625       & 1       & 0.004      & 0.048      & 0.999       & 0.966       \\
\textbf{II} & 0.398 & 0.457   & 0.339    & -1.613       & 0.074       & 0.015       & 0.032      & 0.079      & 0.895       & 0.880       \\ \bottomrule
\end{tabular}
\end{table}

\begin{figure}[H]
	\centering
    \includegraphics{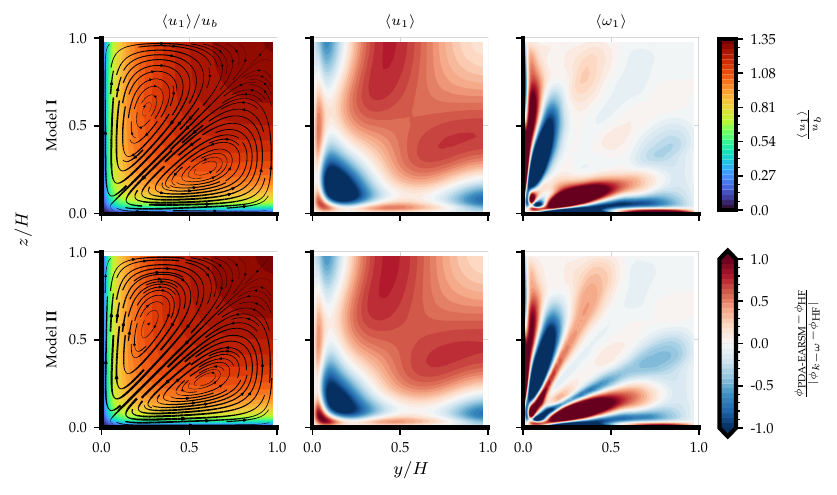}
	\caption{Qualitative streamwise and secondary flow prediction of the best cases for each of the models. Left: streamwise flow contours with stream function lines depicting secondary flow prediction and direction. Centre: streamwise flow error compared to DNS data. Right: streamwise vorticity error compared to DNS data. It should be noted that the stream function line thickness denotes the secondary flow intensity. In the colourbar, $\phi$ indicates either $\langle u_{1} \rangle$ or $\langle \omega_{1} \rangle$.}
	\label{fig:Fig10}
\end{figure}
Regarding qualitative results for the optimisation case, both models are able to accurately predict the direction and symmetry of the secondary motions while improving the streamwise flow prediction. However, the streamwise vorticity error is generally higher in the near-wall regions (Fig. \ref{fig:Fig10}). As a consequence of the prediction of a correct secondary flow direction, the streamwise flow prediction likewise improves. Regions of slight overprediction of $\langle u_{1} \rangle$ are seen in the near-wall vicinity as well as the middle section of the computational volume, whereas regions of $\langle u_{1} \rangle$ underprediction are seen in the bulk flow close to the channel vertex. Regarding the prediction of $\langle \omega_{1} \rangle$, anti-symmetric predictions are seen with respect to the diagonal symmetry line with a general over and underprediction in the near-wall region.

\begin{figure}[H]
	\centering
    \includegraphics{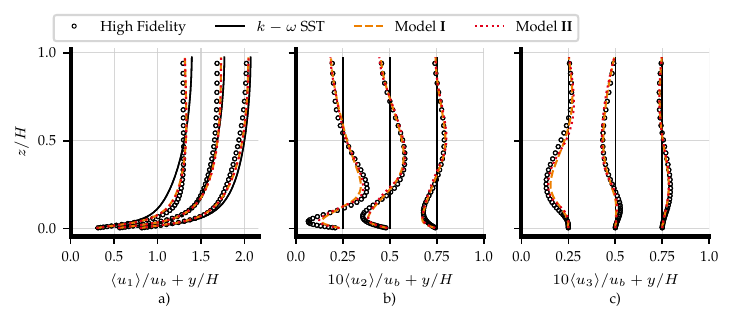}
	\caption{Profiles of velocity components for duct flow case with $\text{AR}=1$ and Re$_{b}=3500$. High-fidelity data obtained from Ref.~\cite{pinelli2010reynolds}.}
	\label{fig:Fig11}
\end{figure}

As expected from the objective function results, qualitative differences between models (Fig. \ref{fig:Fig11}) are not clearly seen. For the streamwise velocity prediction, no significant differences can be seen in the contours of the error function, whereas for the vorticity prediction, slight variations in the error can be seen between models without a significant impact on the overall performance of the models. In summary, both models are able to predict the secondary flow with high accuracy.

Concerning the quantitative analysis of the velocity profiles at $y/H = [0.25, 0.5, 0.75]$, an overall and significant improvement in the prediction of both models against standard $k-\omega$ SST, can be seen. Although the prediction of both models displays minor differences between each other, they both predict with high accuracy the DNS data. Some light discrepancy can be seen in the magnitudes at near-wall regions, where the gradient of $\langle u_{2} \rangle$ is high. Nevertheless, gradients of velocity are accurately predicted, only displaying a slight underprediction in the velocity magnitude in near-wall regions.

\begin{figure}[H]
	\centering
	\includegraphics{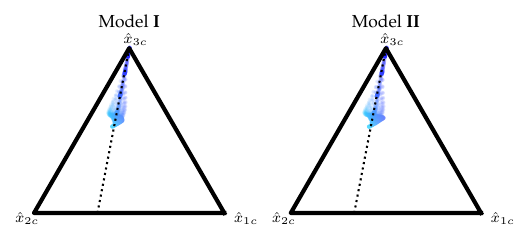}
	\caption{Barycentric map showing the physical reliasability and turbulence anisotropy of the developed models following the colour representation by \cite{emory2014visualizing}.}
	\label{fig:Fig12}
\end{figure}

\begin{figure}[H]
	\centering
        \includegraphics[width=1\textwidth]{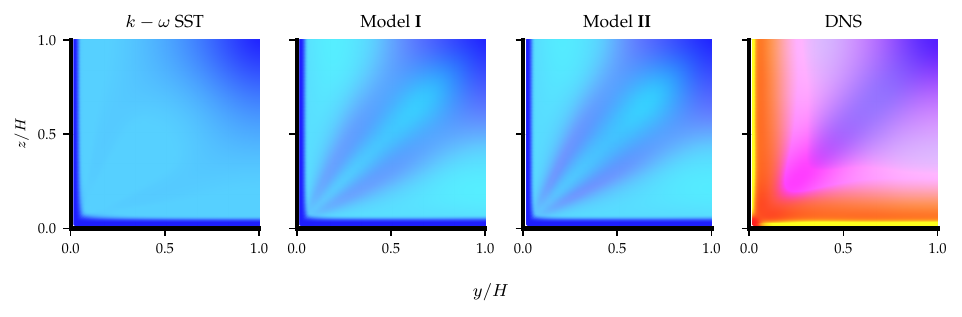}
	\caption{Contours of RSTs' shape for duct flow case with $\text{AR}=1$ and Re$_{b}=3500$. The colours follow the reference barycentric map in Fig.\ref{fig:Fig4}. DNS data from Ref.~\cite{pinelli2010reynolds}.}
	\label{fig:Fig13}
\end{figure}

The turbulence shape of the developed models is represented in Figs.~\ref{fig:Fig12} and ~\ref{fig:Fig13}, where the reliasability for both models is conserved. A certain level of anisotropy is added to the models, yielding results in the vicinity of the plain-strain limit. Both models predict a very similar turbulence shape while model \textbf{II} slightly yields results farther away from the plain-strain limit. Nonetheless, results are distant from $\hat{x}_{1_{c}}$ and DNS anisotropy. 
% This highlights the limitations of the integration of $T_{ij}^{(2)}$ on RANS $k-\omega$ SST, which, while the model is robust, imposes stability limits.
Although these results are expected, the integration of the whole Pope's decomposition with its 10 tensors and a more thorough definition of $\mathcal{B}$ could be able to further improve the RANS anisotropy prediction.

\begin{figure}[H]
	\centering
    \includegraphics{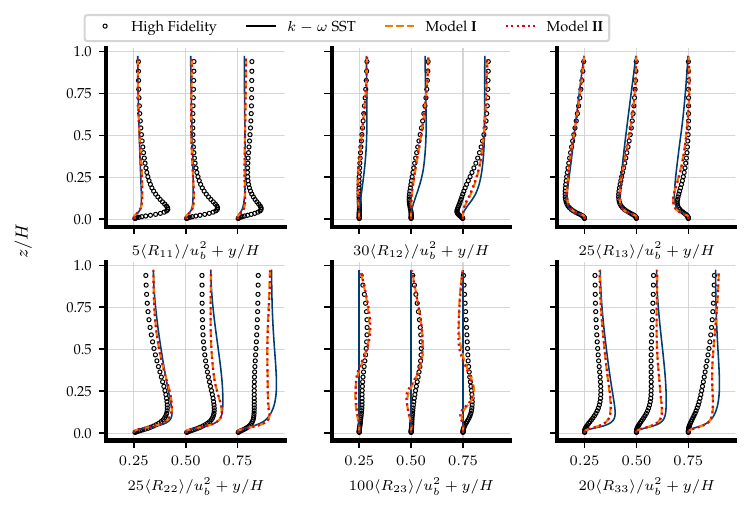}
	\caption{Profiles of Reynolds stress components for duct flow case with $\text{AR}=1$ and Re$_{b}=3500$. High-fidelity data obtained from Ref.~\cite{pinelli2010reynolds}.}
	\label{fig:Fig14}
\end{figure}

Figure~\ref{fig:Fig14} presents the profiles of RST components obtained by new models and shows that new correction models improve the prediction of RST components. It should be mentioned that the Reynolds stress tensor is calculated as 
\begin{equation}
    R_{ij} = A_{ij} + \frac{2}{3}k\delta_{ij}.
\end{equation}

The results of $R_{ij}$ show that the introduced correction does not change the original $R_{11}$ prediction by $k-\omega$ SST while the 2 other principal components of the tensor ($R_{22}$ and $R_{33}$) are predicted in better agreement with high-fidelity data. Predictions of the off-diagonal components of $R_{ij}$ are overall displaying an improvement in prediction, highlighting the non-zero prediction of $R_{23}$, which shows a certain level of mismatch with high-fidelity data in the near-wall regions.

Qualitatively and quantitatively, both models improve the prediction and agreement with high-fidelity Reynolds stresses while not showing significant differences in the prediction of Reynolds stresses between models. 

\subsection{Verification and generalisability on test cases}
\label{sec:verficationAndGeneralisability}
In this subsection, the best-found models are tested against diverse canonical cases in which secondary flow is an important component as well as cases where secondary flow is not present. This verification is performed in order to provide a performance overview of the models' generalisability, stability, and robustness to preserve canonical flow features. Firstly, validation of the channel case and the boundary layer is performed at different Reynolds numbers. Secondly, the models are tested against diverse cases of ducts with different aspect ratios and Reynolds numbers. Lastly, the models are tested against a wall-modelled nominally infinite Reynolds number case in which secondary flow is roughness-induced. Similarly as in section \ref{sec:results}, all results are shown and discussed with their respective velocity contour plots with stream functions and qualitative flow analysis of the velocity profiles.

\subsubsection{Channel flow and law of the wall}
The addition of $T_{ij}^{(2)}$ and further modifications in this study must not destabilise $k-\omega$ SST and yield unphysical results. Therefore, both models are tested in a channel flow case at friction Reynolds numbers 395 and 5200 to verify that the law of the wall is preserved. The canonical channel flow cases are driven by a constant pressure gradient in order to match the friction Reynolds number from DNS.

\begin{figure}[H]
     \centering
     \includegraphics{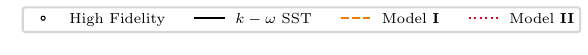}
    \\
     \begin{subfigure}[t]{0.495\linewidth}
        \centering
        \includegraphics[width=1\textwidth]{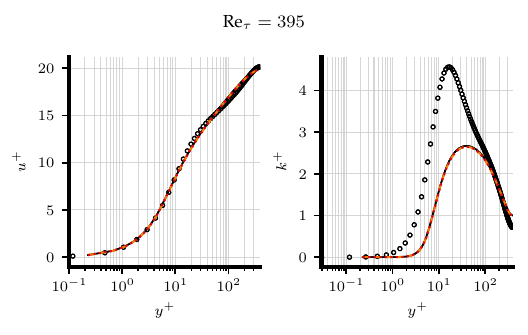}
        \caption{}
        \label{fig:Fig15a}
     \end{subfigure}
     \hfill
     \begin{subfigure}[t]{0.495\linewidth}
        \centering
        \includegraphics[width=1\textwidth]{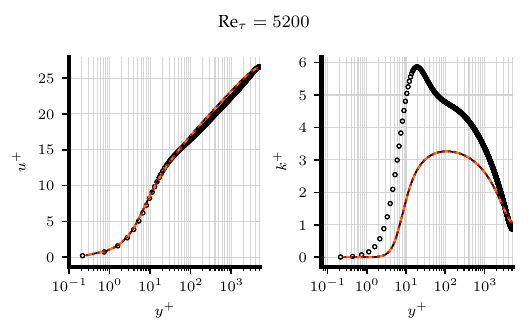}
        \caption{}
        \label{fig:Fig15b}
     \end{subfigure}
     \caption{Mean streamwise velocity and turbulent kinetic energy profiles for channel flow: $\text{Re}_{\tau}=395$ (Fig. \ref{fig:Fig15a}), and Re$_{\tau}=5200$ (Fig. \ref{fig:Fig15b}). High-fidelity data obtained from Ref.~\cite{moser1999direct}.}
     \label{fig:Fig15}
\end{figure}

Results shown in Fig. \ref{fig:Fig15} depict the streamwise velocity as $u^{+} = u/u_{\tau}$ and $k^{+}=k/u_{\tau}^{2}$ in function of $y^{+}=u_{\tau}y/\nu$, where $u_{\tau}$ is the friction velocity. The results show a consistent prediction of the law of the wall with $k-\omega$ SST without introducing any noticeable changes. Both models yield the same results for both the streamwise velocity and the turbulence kinetic energy, and no improvements or diminishments in the prediction of these variables are seen compared to standard $k-\omega$ SST.

% \red

% \subsubsection{Separated flows}
% In order to further validate the models, the periodic hill canonical flow case at $\text{Re}_{b} = 2800$ is evaluated. This case displays boundary-layer separation and reattachment with no secondary flow, therefore, it is expected that the developed models do not change the $k-\omega$ SST prediction.

% Figure \ref{fig:PH2800} displays the quantitative predictions of DNS, $k-\omega$ SST, and the developed models for this case, where no visible changes are seen compared to the standard prediction of $k-\omega$ SST for model \textbf{II}. However, model \textbf{I} yields numerical instabilities that prevent the solution from achieving convergence, generating a noisy prediction in both the vertical velocity and the friction coefficient.
% \begin{figure}[H]
% 	\centering
% 	\includegraphics[width=1\textwidth]{PH2800.pdf}
% 	\caption{\red Periodic hill velocity profiles (left) and friction coefficient (right) for $\text{Re}_{b}=2800$. High-fidelity data obtained from Ref.~\cite{balakumar2015dns}. \black}
% 	\label{fig:PH2800}
% \end{figure}

% Despite the general agreement and consistency of both models on the training case and channel flow, only model \textbf{II} matches the original behaviour of $k-\omega$ SST while preserving numerical stability in this case.

% \black

\subsubsection{Duct cases}
To showcase the generalisability of the models, diverse duct cases are tested. The first case is the flow through a duct of AR $=1$ and a considerably higher Re$_{b}=5700$ compared to the optimisation case.

\begin{figure}[H]
	\centering
	\includegraphics[width=1\textwidth]{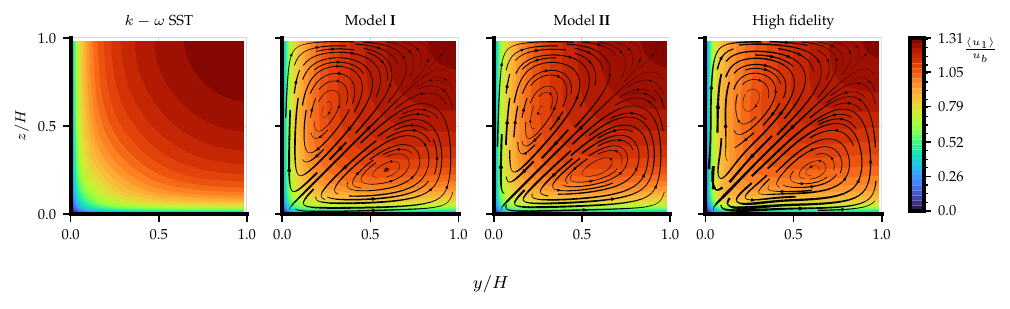}
	\caption{Streamwise flow contours with stream function iso-lines depicting secondary flow prediction and direction: Duct flow case with $\text{AR}=1$ and Re$_{b}=5700$. High-fidelity data obtained from Ref.~\cite{vinuesa2018secondary}.}
	\label{fig:Fig16}
\end{figure}

A qualitative depiction of the results is shown in Fig. \ref{fig:Fig16}, where both models are able to predict the gradients of the secondary flow and improve the prediction of the streamwise velocity compared to $k-\omega$ SST. In agreement with the baseline results of the models, the magnitude of the secondary motion is weaker than the high-fidelity data although the motion's antisymmetry and location of the vortices centres are predicted accurately.

\begin{figure}[H]
	\centering
	\includegraphics{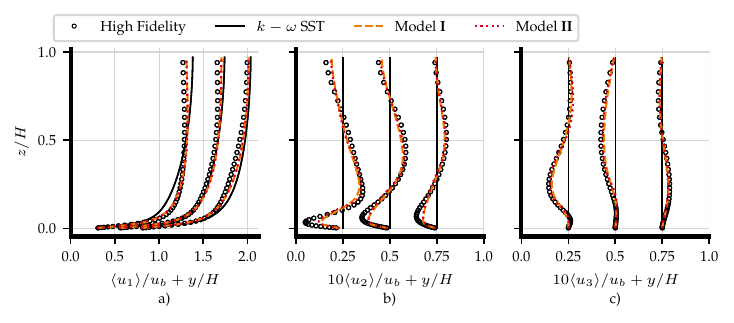}
	\caption{Profiles of velocity components for duct flow case with $\text{AR}=1$ and Re$_{b}=5700$. High-fidelity data obtained from Ref.~\cite{vinuesa2018secondary}.}
	\label{fig:Fig17}
\end{figure}

Regarding the velocity profiles of the case (Fig. \ref{fig:Fig17}), a similar trend is observed where the gradients and magnitude of the profiles are predicted with high accuracy in both models, only displaying some discrepancies in the wall-near regions, where the high-fidelity data yields slightly higher velocity gradients.

In order to verify the performance of the models in diverse cases with similar features, the models are likewise tested against a duct of AR $=3$ and a slightly lower Re$_{b}=2600$ compared to the optimisation case.

\begin{figure}[H]
	\centering
	\includegraphics{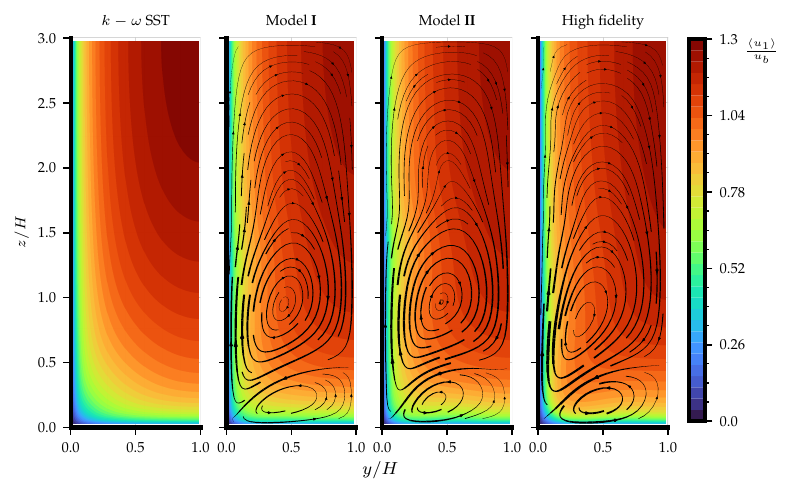}
	\caption{Streamwise flow contours with stream function iso-lines depicting secondary flow prediction and direction: Duct flow case with $\text{AR}=3$ and Re$_{b}=2600$. High-fidelity data obtained from Ref.~\cite{vinuesa2018secondary}.}
	\label{fig:Fig18}
\end{figure}

The qualitative results shown in Fig. \ref{fig:Fig18} indicate a similar performance for both models: gradients, symmetry, and direction of secondary flow are predicted accurately with slight inaccuracies regarding the secondary flow magnitude in the near-wall regions. The location of the vortices centres is likewise predicted with high accuracy, agreeing with high-fidelity data (Fig. \ref{fig:Fig18}).

\begin{figure}[H]
	\centering
	\includegraphics{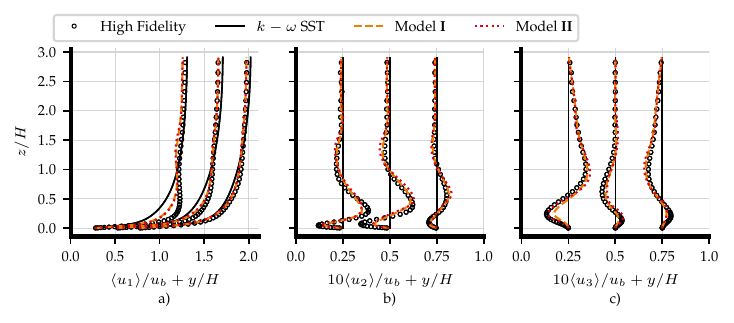}
	\caption{Profiles of velocity components for duct flow case with $\text{AR}=3$ and Re$_{b}=2600$. High-fidelity data obtained from Ref.~\cite{vinuesa2018secondary}.}
	\label{fig:Fig19}
\end{figure}

In terms of the quantitative analysis and the evaluation of the velocity profiles, the results shown in Fig. \ref{fig:Fig19} likewise follow a similar trend compared to previous results in this study. The region in the vicinity of the duct vertex shows some discrepancy in the velocity prediction. Nevertheless, the bulk flow away from the near wall is accurately predicted.

\subsubsection{Roughness-induced secondary flow case}
Finally, the verification of the models is tested on a roughness-induced secondary flow case. The case is based on the studies by \cite{forooghi2020roughness, amarloo2022secondary}, and is characterised by displaying a nominally infinite Reynolds number. It should be noted that due to the very high Reynolds number, the use of wall models is inevitable with current hardware, therefore, the atmospheric wall models \cite{parente2011improved} have been used. The secondary flows generated in this case are recognised as Prandtl's second kind of secondary flow, similar to the square duct flow \cite{nikitin2021prandtl}. The roughness-induced case is chosen to showcase the models' performance in a more complex and challenging to predict case, mostly by two-equation RANS turbulence models.
% Furthermore, this case is likewise known for the difficulties of two-equation RANS turbulence models to accurately predict.

For a better analysis of roughness-induced secondary flow, dispersive velocity components are defined as,
\begin{equation}
    \langle u_{i}^{\prime \prime} \rangle = \langle u_{i} \rangle - \langle \tilde{u}_{i} \rangle,
\end{equation}
where $\langle u_{i}^{\prime \prime} \rangle$ is the dispersive velocity components, and $\langle \tilde{u}_{i} \rangle$ is the spatial spanwise-averaged mean velocity.

\begin{figure}[H]
	\centering
	\includegraphics{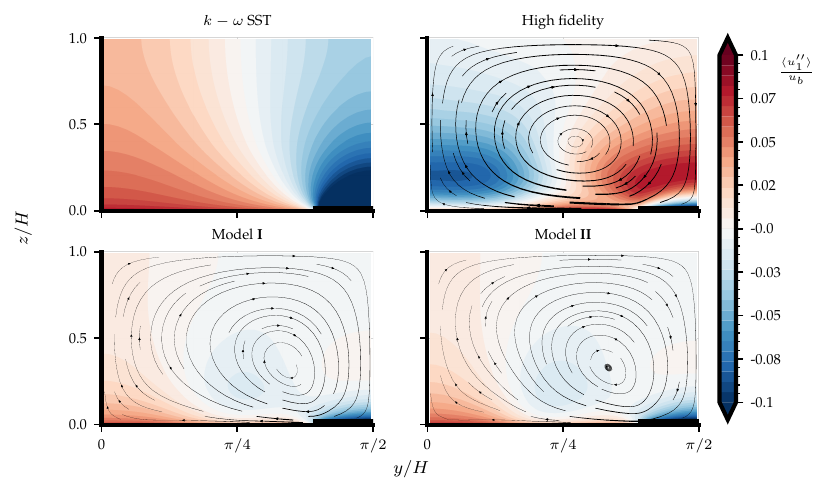}
	\caption{Streamwise dispersive velocity contours with stream function iso-lines depicting secondary flow prediction and direction: Roughness-induced secondary flow case with Re$_{b}=2\times10^{8}$. High-fidelity data obtained from Ref.~\cite{amarloo2022secondary}.}
	\label{fig:Fig20}
\end{figure}

Following previous verification, the qualitative results of the streamwise dispersive velocity ($\langle u_{1}^{\prime \prime} \rangle$) are shown in Fig. \ref{fig:Fig20}. On the one hand, results show that standard $k-\omega$ SST does not predict any secondary motion that drives the flow to display a high-momentum path on top of the high-roughness patch and a low-momentum path at $y/H=0$, as the high-fidelity data predicts. On the other hand, the enhanced models are both capable of predicting the secondary motion, although at a lower intensity compared to high-fidelity results. The direction of the rotation (clockwise) is correctly predicted although the vortex centre does not visibly match the high-fidelity counterpart. Even though the predicted secondary flow is not strong enough to move the low-momentum path to the top of the low-roughness patch ($y/H=0$), the existence of a weak high-momentum path is visible on top of the high-roughness patch. Since the source of vorticity production lies at roughness-heterogeneity existing at the bottom wall (more information at Ref.~\cite{amarloo2022secondary}), the authors believe that further investigations about wall models regarding the secondary flow can help a better prediction of the secondary flow.

\begin{figure}[H]
	\centering
	\includegraphics{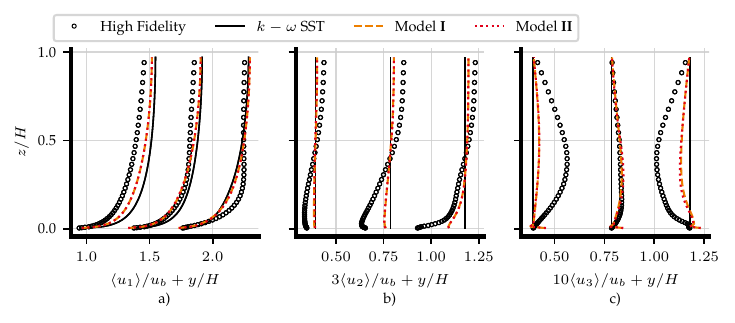}
	\caption{Profiles of velocity components for roughness-induced secondary flow with Re$_{b}=2\times10^{8}$. High-fidelity data obtained from Ref.~\cite{amarloo2022secondary}.}
	\label{fig:Fig21}
\end{figure}

\begin{figure}[H]
	\centering
	\includegraphics{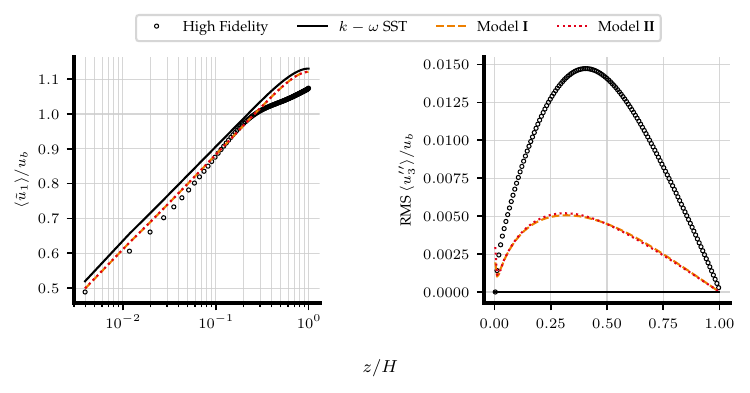}
	\caption{Roughness-induced secondary flow quantitative results of spanwise-averaged streamwise velocity (left), and root-mean-squared dispersive vertical velocity (right). High-fidelity data obtained from Ref.~\cite{amarloo2022secondary}.}
    \label{fig:Fig22}
\end{figure}

Quantitive results reflect the previous qualitative analysis. Figure~\ref{fig:Fig21} shows the velocity profiles at $y/H = \left[ \pi/8, \pi/4, 3\pi/8 \right]$, where the improvement of the models can be seen. In contrast to other verification cases in this study, both models' predictions improve the performance of standard $k-\omega$ SST, however, the secondary flow intensity is weaker compared to the high-fidelity data. The predicted tendencies are correct, although there is a discrepancy in the gradients and magnitudes predictions relative to previous verification cases.
To provide a complete and holistic comparative study for this case, the spatial average of the streamwise, as well as vertical root-mean-squared (RMS) dispersive velocity ($\langle \tilde{u}_{1} \rangle$ and RMS $\langle u_{3}^{\prime \prime} \rangle$, respectively), are calculated. These results are shown in Fig. \ref{fig:Fig22}, where the prediction of $\langle \tilde{u}_{1} \rangle$ shows a considerable improvement compared to standard $k-\omega$ SST. However, the bulk flow for the developed models is still overpredicted as a consequence of the underpredicted secondary flow and its derived under-subtraction of the streamwise momentum. Regarding the RMS $\langle u_{3}^{\prime \prime} \rangle$, a clear improvement in the prediction can be seen by the developed models, where $k-\omega$ SST is unable to predict any vertical dispersive velocity. The predictive profiles are underpredicted and slightly skewed towards $z/H=0$ compared to the high-fidelity data, which verifies the mismatched location of the predicted secondary flow vortices. Both developed models yield an almost identical prediction of the flow, and a similar improvement is seen by models at their prediction of RMS $\langle u_{3}^{\prime \prime} \rangle$. Overall, the applicability of the models to this case shows their generalisability, where robust and improved predictions are obtained in more complex cases at very high Reynolds numbers.
\section{Comparison with other EARSMs}
\blue To provide a complete and updated performance of the developed models, a comparison is made against the two novel machine-learned EARSMs developed by Saidi et al. \cite{saidi2022cfd} denoted as M1 (Model 1) and M2 (Model 2); and the classical (not machined-learned) EARSM BSL-EARSM based on $k-\omega$ SST model and on the explicit constitutive relation by Ref. \cite{wallin2000explicit} and developed by Menter et al. \cite{menter2012explicit}. The comparison is made on CF$_{395}$, CF$_{5200}$, and duct flow with Re$_{b}=10320$ and AR$=1$.

\begin{figure}[H]
     \centering
     \includegraphics{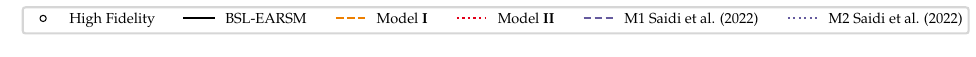}
    \\
     \begin{subfigure}[t]{0.495\linewidth}
        \centering
        \includegraphics{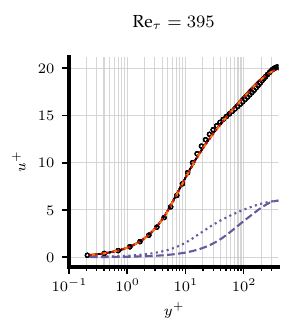}
        \caption{}
        \label{fig:Fig23a}
     \end{subfigure}
     \hfill
     \begin{subfigure}[t]{0.495\linewidth}
        \centering
        \includegraphics{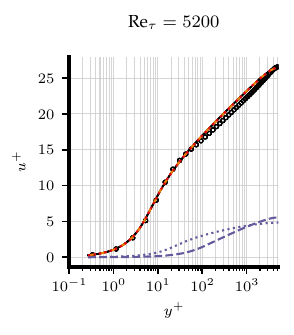}
        \caption{}
        \label{fig:Fig23b}
     \end{subfigure}
     \caption{Model comparison of mean streamwise velocity and turbulent kinetic energy profiles for channel flow: $\text{Re}_{\tau}=395$ (Fig. \ref{fig:Fig23a}), and Re$_{\tau}=5200$ (Fig. \ref{fig:Fig23b}). High-fidelity data obtained from Ref.~\cite{moser1999direct}.}
     \label{fig:Fig23}
\end{figure}

As shown in the comparative results of Fig. \ref{fig:Fig23}, Saidi et al. models mispredict the prediction of boundary layers due to its modification of $T_{ij}^{(1)}$ in their models. These modifications improve the prediction of separated flows at the expense of modifying the successful prediction of the law of the wall by standard $k-\omega$ SST. This undesirable effect, however, can be avoided by deactivating the $T_{ij}^{(1)}$ correction in the regions corresponding to equilibrium boundary layers (e.g., a space-dependant aggregation in Ref.~\cite{cherroud2023space}). Since the work performed in this study only adds the contributions of $T_{ij}^{(2)}$, the developed models are able to preserve the successful predictions of $k-\omega$ SST for equilibrium boundary layers.

\begin{figure}[H]
	\centering
	\includegraphics{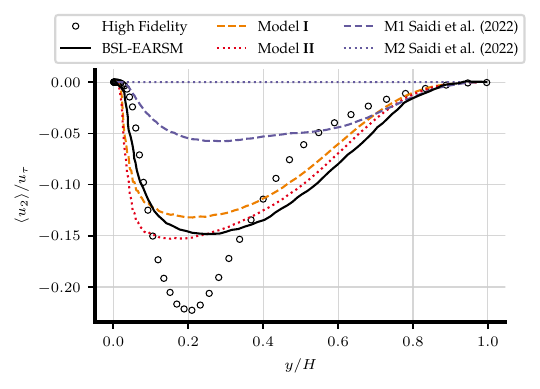}
	\caption{Normalised vertical velocity component comparison between models for DF$_{10320}$ with AR$=1$ along line $y=z$. High-fidelity data obtained from Ref.~\cite{huser1993direct}.}
    \label{fig:Fig24}
\end{figure}

Finally, a comparison showing the prediction of the $\langle u_{2} \rangle$ component along the $y=z$ line of a square duct at Re$_{b}=10320$ is shown in Fig. \ref{fig:Fig24}. It can be seen that the PDA-EARSMs outperform the models by Saidi et al. and yield similar performance compared to BSL-EARSM.
\black
\section{Conclusions}
\label{sec:conclusions}
This study employs CFD-driven surrogate and Bayesian optimisation to enhance the $k-\omega$ SST turbulence model focusing on predicting Prandtl's second kind of secondary flow. A progressive approach is adopted to enhance the linear eddy-viscosity model with the ability to predict secondary flows while maintaining its effective performance in canonical flow cases. The enhancement is based on the introduction of explicit algebraic Reynolds stress correction models with two levels of complexity: introducing a linear combination of the first two candidate functions and introducing a linear combination of the first two principal components obtained from PCA on all candidate functions. These two models are optimised to reach the best prediction of both secondary flow and streamwise flow in the optimisation case of a duct flow with an aspect ratio of 1 and $\text{Re}_b=3500$.

Considering the progressive approach, the enhanced models are tested against channel flow cases at different friction Reynolds numbers, where the enhanced models preserve the original performance of the $k-\omega$ SST model in a successful reproduction of the velocity and TKE profiles. 

For the purpose of generalisability investigation, new models are tested against verification duct-flow cases with different Reynolds numbers and different aspect ratios. Both enhanced models show a successful prediction of secondary flow and improvement of streamwise velocity prediction in the unseen cases. The third and final verification case is a roughness-induced secondary flow case at a nominally infinite Reynolds number, where the enhanced models are able to predict the secondary motion, although at a lower intensity compared to high-fidelity data due to the limitations of wall modelling for this type of cases.

Overall, the results, summarised in Table~\ref{tab:Table2}, show that the developed models perform substantially better than the standard $k-\omega$ SST model in all verification cases. The enhanced models are able to predict secondary flow features, showing global improvements in the velocity field prediction. Both models show similar performance in successfully predicting secondary flows and streamwise velocity. Therefore, more invariants combined with PCA do not necessarily improve the performance of the models in the 2-dimensional cases considered in this study. Nevertheless, considering scaling up the complexity of the model in 3-dimensional cases, the combination with PCA dimensionality reduction could further improve the predictions. It should be noted that the optimisation process and the nature of the PDA-EARSMs yield a certain level of turbulence anisotropy while always maintaining solution robustness and stability. This fact results in some discrepancies between the PDA-EARSMs and high-fidelity data in the predicted turbulence shape.

\begin{table}[H]
\centering
\caption{Objective function results for optimisation case and all verification cases tested. Results are presented by the case abbreviation and characteristics: Duct flow cases as DF$_{\text{AR}, \ \text{Re}}$ and roughness-induced secondary flow as RI$_{\text{Re}}$. It should be noted that DF$_{1, \ 3500}$ is the optimisation case. $J=1$ indicates the $k-\omega$ SST original prediction and $J=0$ indicates a match of the high-fidelity data.}
\label{tab:Table2}
\begin{tabular}{@{}lllll@{}}
\toprule
Case                                                                            & Model       & $J$   & $j_{1}$ & $j_{2}$ \\ \midrule
\multirow{2}{*}{\begin{tabular}[c]{@{}l@{}}DF$_{1, \ 3500}$\end{tabular}} & \textbf{I}  & 0.387 & 0.455   & 0.319   \\
                                                                                & \textbf{II} & 0.398 & 0.457   & 0.339   \\ \midrule
\multirow{2}{*}{\begin{tabular}[c]{@{}l@{}}DF$_{1, \ 5700}$\end{tabular}} & \textbf{I}  & 0.405 & 0.482   & 0.328   \\
                                                                                & \textbf{II} & 0.392 & 0.461   & 0.323   \\ \midrule
\multirow{2}{*}{\begin{tabular}[c]{@{}l@{}}DF$_{3, \ 2600}$\end{tabular}} & \textbf{I}  & 0.458 & 0.551   & 0.365   \\
                                                                                & \textbf{II} & 0.502 & 0.580   & 0.424   \\ \midrule
\multirow{2}{*}{\begin{tabular}[c]{@{}l@{}}RI$_{2\times10^{8}}$\end{tabular}}   & \textbf{I}  & 0.693 & 0.663   & 0.723   \\
                                                                                & \textbf{II} & 0.698 & 0.666   & 0.730   \\ \midrule
\multirow{2}{*}{\begin{tabular}[c]{@{}l@{}}Average\end{tabular}}   & \textbf{I}  & \textbf{0.486} & \textbf{0.538}   & \textbf{0.434}   \\
                                                                                    & \textbf{II} & \textbf{0.498} & \textbf{0.541}   & \textbf{0.454}   \\ \bottomrule
% \begin{tabular}[c]{@{}l@{}}Total Average\end{tabular}   &   & \textbf{0.492} & \textbf{0.540}   & \textbf{0.444}   \\ \bottomrule
\end{tabular}
\end{table}

These findings suggest that the use of CFD-driven optimisation with surrogate modelling and Bayesian optimisation is a robust approach to enhance linear-eddy-viscosity turbulence models to predict more complex physics. The PDA-EARSMs developed greatly improve the prediction of turbulence anisotropy in wall-bounded flows, especially in cases where the standard $k-\omega$ SST presents difficulties in accurately predicting the secondary flow. The progressive nature of this development as a first step focuses on the prediction of secondary flows, allowing further improvement of the models by adding more predictive physics and verifying them on other test cases.

\section*{CRediT authorship contribution statement}
\textbf{Mario Javier Rinc\'on}: Conceptualisation, Methodology, Software, Validation, Formal analysis, Writing – original draft. 
\textbf{Ali Amarloo}: Conceptualisation, Methodology, Software, Validation, Formal analysis, Writing – original draft.
\textbf{Martino Reclari}: Formal analysis, Methodology, Writing – review \& editing. 
\textbf{Xiang I. A. Yang}: Formal analysis, Methodology, Writing – review \& editing. 
\textbf{Mahdi Abkar}: Conceptualisation, Methodology, Formal analysis, Project administration, Resources, Supervision, Writing – review \& editing.

\section*{Declaration of Competing Interest}
The authors declare that they have no known competing financial interests or personal relationships that could have appeared to influence the work reported in this study.

\section*{Acknowledgments}
M.J.R. acknowledges the financial support from the Innovation Fund Denmark (IFD) under Grant No. 0153-00051B and Kamstrup A/S. A.A. acknowledges the financial support from the Aarhus University Centre for Digitalisation, Big Data, and Data Analytics (DIGIT) and the Aarhus University Research Foundation (AUFF). 
%This research is supported by  The authors also acknowledge the financial support from the Aarhus University Centre for Digitalisation, Big Data, and Data Analytics (DIGIT) and Aarhus University Research Foundation (AUFF). 
This work was also partially supported by the Danish e-Infrastructure Cooperation (DeiC) National HPC under grant numbers DeiC-AU-N2-2023003 and DeiC-AU-N5-2023004.

\section*{Data availability}
The models developed in this study are publicly available for their use and compilation in OpenFOAM under \href{https://github.com/AUfluids/PDA2DEARSM}{https://github.com/AUfluids/PDA2DEARSM}. Further data will be made available on request.

% \appendix
% \section{Appendix A}


%\bibliographystyle{cas-model2-names}
\bibliographystyle{elsarticle-num} 

\bibliography{cas-refs}

\begin{thebibliography}{10}
\expandafter\ifx\csname url\endcsname\relax
  \def\url#1{\texttt{#1}}\fi
\expandafter\ifx\csname urlprefix\endcsname\relax\def\urlprefix{URL }\fi
\expandafter\ifx\csname href\endcsname\relax
  \def\href#1#2{#2} \def\path#1{#1}\fi

\bibitem{slotnick2014cfd}
J.~Slotnick, A.~Khodadoust, A.~Juan, D.~Darmofal, W.~Gropp, E.~Lurie,
  D.~Mavriplis, {CFD} vision 2030 study: A path to revolutionary computational
  aerosciences, {Technical Report} NASA/CR-2014-218178, NASA (2014).

\bibitem{nikitin2021prandtl}
N.~Nikitin, N.~Popelenskaya, A.~Stroh, Prandtl’s secondary flows of the
  second kind. {P}roblems of description, prediction, and simulation, Fluid
  Dyn. 56~(4) (2021) 513--538.

\bibitem{xiao2019quantification}
H.~Xiao, P.~Cinnella, Quantification of model uncertainty in {RANS}
  simulations: A review, Prog. Aerosp. Sci. 108 (2019) 1--31.

\bibitem{duraisamy2019turbulence}
K.~Duraisamy, G.~Iaccarino, H.~Xiao, Turbulence modeling in the age of data,
  Annu. Rev. Fluid Mech. 51 (2019) 357--377.

\bibitem{tracey2013application}
B.~Tracey, K.~Duraisamy, J.~Alonso, Application of supervised learning to
  quantify uncertainties in turbulence and combustion modeling, in: 51st AIAA
  aerospace sciences meeting including the new horizons forum and aerospace
  exposition, 2013, p. 259.

\bibitem{wang2017physics}
J.-X. Wang, J.-L. Wu, H.~Xiao, Physics-informed machine learning approach for
  reconstructing {R}eynolds stress modeling discrepancies based on {DNS} data,
  Phys. Rev. Fluids 2~(3) (2017) 034603.

\bibitem{wu2018physics}
J.-L. Wu, H.~Xiao, E.~Paterson, Physics-informed machine learning approach for
  augmenting turbulence models: A comprehensive framework, Phys. Rev. Fluids
  3~(7) (2018) 074602.

\bibitem{ling2016reynolds}
J.~Ling, A.~Kurzawski, J.~Templeton, {R}eynolds averaged turbulence modelling
  using deep neural networks with embedded invariance, J. Fluid Mech. 807
  (2016) 155--166.

\bibitem{kaandorp2020data}
M.~L. Kaandorp, R.~P. Dwight, Data-driven modelling of the {R}eynolds stress
  tensor using random forests with invariance, Comput. Fluids 202 (2020)
  104497.

\bibitem{mcconkey2022deep}
R.~McConkey, E.~Yee, F.-S. Lien, Deep structured neural networks for turbulence
  closure modeling, Phys. Fluids 34~(3) (2022) 035110.

\bibitem{cruz2019use}
M.~A. Cruz, R.~L. Thompson, L.~E. Sampaio, R.~D. Bacchi, The use of the
  {R}eynolds force vector in a physics informed machine learning approach for
  predictive turbulence modeling, Comput. Fluids 192 (2019) 104258.

\bibitem{weatheritt2016novel}
J.~Weatheritt, R.~Sandberg, A novel evolutionary algorithm applied to algebraic
  modifications of the {RANS} stress--strain relationship, J. Comput. Phys. 325
  (2016) 22--37.

\bibitem{weatheritt2017development}
J.~Weatheritt, R.~Sandberg, The development of algebraic stress models using a
  novel evolutionary algorithm, Int. J. Heat Fluid Flow 68 (2017) 298--318.

\bibitem{schmelzer2020discovery}
M.~Schmelzer, R.~P. Dwight, P.~Cinnella, Discovery of algebraic
  {R}eynolds-stress models using sparse symbolic regression, Flow Turbul.
  Combust. 104~(2) (2020) 579--603.

\bibitem{amarloo2022frozen}
A.~Amarloo, P.~Forooghi, M.~Abkar, Frozen propagation of {R}eynolds force
  vector from high-fidelity data into {R}eynolds-averaged simulations of
  secondary flows, Phys. Fluids 34~(11) (2022) 115102.

\bibitem{duraisamy2021perspectives}
K.~Duraisamy, Perspectives on machine learning-augmented {R}eynolds-averaged
  and large eddy simulation models of turbulence, Phys. Rev. Fluids 6~(5)
  (2021) 050504.

\bibitem{singh2017machine}
A.~P. Singh, S.~Medida, K.~Duraisamy, Machine-learning-augmented predictive
  modeling of turbulent separated flows over airfoils, AIAA J. 55~(7) (2017)
  2215--2227.

\bibitem{holland2019field}
J.~R. Holland, J.~D. Baeder, K.~Duraisamy, Field inversion and machine learning
  with embedded neural networks: Physics-consistent neural network training,
  in: AIAA Aviation 2019 Forum, 2019, p. 3200.

\bibitem{zhao2020rans}
Y.~Zhao, H.~D. Akolekar, J.~Weatheritt, V.~Michelassi, R.~D. Sandberg, {RANS}
  turbulence model development using {CFD}-driven machine learning, J. Comput.
  Phys. 411 (2020) 109413.

\bibitem{saidi2022cfd}
I.~B.~H. Sa{\"\i}di, M.~Schmelzer, P.~Cinnella, F.~Grasso, {CFD}-driven
  symbolic identification of algebraic {R}eynolds-stress models, J. Comput.
  Phys. 457 (2022) 111037.

\bibitem{barzilai1988two}
J.~Barzilai, J.~M. Borwein, Two-point step size gradient methods, IMA J. Numer.
  Anal. 8~(1) (1988) 141--148.

\bibitem{nelder1965simplex}
J.~A. Nelder, R.~Mead, {A Simplex Method for Function Minimization}, Comput. J.
  7~(4) (1965) 308--313.

\bibitem{coello2007evolutionary}
C.~A.~C. Coello, G.~B. Lamont, D.~A. Van~Veldhuizen, et~al., Evolutionary
  algorithms for solving multi-objective problems, Vol.~5, Springer, (2007).

\bibitem{eberhart1995particle}
R.~Eberhart, J.~Kennedy, Particle swarm optimization, in: Proceedings of the
  {IEEE} international conference on neural networks, Vol.~4, Citeseer, 1995,
  pp. 1942--1948.

\bibitem{forrester2008engineering}
A.~Forrester, A.~Sobester, A.~Keane, Engineering design via surrogate
  modelling: a practical guide, John Wiley \& Sons, 2008.

\bibitem{rincon2022turbulent}
M.~J. Rinc{\'o}n, M.~Reclari, M.~Abkar, Turbulent flow in small-diameter
  ultrasonic flow meters: A numerical and experimental study, Flow Meas.
  Instrum. 87 (2022) 102227.

\bibitem{rincon2023validating}
M.~J. Rinc{\'o}n, M.~Reclari, X.~I. Yang, M.~Abkar, Validating the design
  optimisation of ultrasonic flow meters using computational fluid dynamics and
  surrogate modelling, Int. J. Heat Fluid Flow 100 (2023) 109112.

\bibitem{singh2017multi}
P.~Singh, I.~Couckuyt, K.~Elsayed, D.~Deschrijver, T.~Dhaene, Multi-objective
  geometry optimization of a gas cyclone using triple-fidelity co-kriging
  surrogate models, J. Optim. Theory Appl. 175~(1) (2017) 172--193.

\bibitem{lam2009coupled}
X.~Lam, Y.~Kim, A.~Hoang, C.~Park, Coupled aerostructural design optimization
  using the kriging model and integrated multiobjective optimization algorithm,
  J. Optim. Theory Appl. 142~(3) (2009) 533--556.

\bibitem{urquhart2020aerodynamic}
M.~Urquhart, M.~Varney, S.~Sebben, M.~Passmore, Aerodynamic drag improvements
  on a square-back vehicle at yaw using a tapered cavity and asymmetric flaps,
  Int. J. Heat Fluid Flow 86 (2020) 108737.

\bibitem{huang2021bayesian}
X.~L. Huang, X.~I. Yang, A bayesian approach to the mean flow in a channel with
  small but arbitrarily directional system rotation, Phys. Fluids 33~(1) (2021)
  015103.

\bibitem{xiang2021neuroevolution}
T.-R. Xiang, X.~Yang, Y.-P. Shi, Neuroevolution-enabled adaptation of the
  jacobi method for poisson’s equation with density discontinuities, Theor.
  App. Mech. Lett. 11~(3) (2021) 100252.

\bibitem{waschkowski2022multi}
F.~Waschkowski, Y.~Zhao, R.~Sandberg, J.~Klewicki, Multi-objective {CFD}-driven
  development of coupled turbulence closure models, J. Comput. Phys. 452 (2022)
  110922.

\bibitem{queipo2005surrogate}
N.~V. Queipo, R.~T. Haftka, W.~Shyy, T.~Goel, R.~Vaidyanathan, P.~K. Tucker,
  Surrogate-based analysis and optimization, Progress in aerospace sciences
  41~(1) (2005) 1--28.

\bibitem{sacks1989designs}
J.~Sacks, S.~B. Schiller, W.~J. Welch, Designs for {Computer Experiments},
  Technometrics 31~(1) (1989) 41--47.

\bibitem{sandberg2022machine}
R.~D. Sandberg, Y.~Zhao, Machine-learning for turbulence and heat-flux model
  development: A review of challenges associated with distinct physical
  phenomena and progress to date, Int. J. Heat Fluid Flow 95 (2022) 108983.

\bibitem{fang2023toward}
Y.~Fang, Y.~Zhao, F.~Waschkowski, A.~S. Ooi, R.~D. Sandberg, Toward more
  general turbulence models via multicase computational-fluid-dynamics-driven
  training, AIAA J. (2023) 1--16.

\bibitem{bin2022progressive}
Y.~Bin, L.~Chen, G.~Huang, X.~I. Yang, Progressive, extrapolative machine
  learning for near-wall turbulence modeling, Phys. Rev. Fluids 7~(8) (2022)
  084610.

\bibitem{menter1994two}
F.~R. Menter, Two-equation eddy-viscosity turbulence models for engineering
  applications, {AIAA} J. 32~(8) (1994) 1598--1605.

\bibitem{pope1975more}
S.~B. Pope, A more general effective-viscosity hypothesis, J. Fluid Mech.
  72~(2) (1975) 331--340.

\bibitem{forooghi2020roughness}
P.~Forooghi, X.~I. Yang, M.~Abkar, Roughness-induced secondary flows in stably
  stratified turbulent boundary layers, Phys. Fluids 32~(10) (2020) 105118.

\bibitem{amarloo2022secondary}
A.~Amarloo, P.~Forooghi, M.~Abkar, Secondary flows in statistically unstable
  turbulent boundary layers with spanwise heterogeneous roughness, Theor. App.
  Mech. Lett. 12~(2) (2022) 100317.

\bibitem{wallin2000explicit}
S.~Wallin, A.~V. Johansson, An explicit algebraic reynolds stress model for
  incompressible and compressible turbulent flows, J. Fluid Mech. 403 (2000)
  89--132.

\bibitem{pope2000turbulent}
S.~B. Pope, Turbulent flows, Cambridge University Press, (2000).

\bibitem{menter2003ten}
F.~R. Menter, M.~Kuntz, R.~Langtry, Ten years of industrial experience with the
  sst turbulence model, Turbul. Heat Mass Transf. 4~(1) (2003) 625--632.

\bibitem{jones2001taxonomy}
D.~R. Jones, A taxonomy of global optimization methods based on response
  surfaces, J. Glob. Optim. 21~(4) (2001) 345--383.

\bibitem{weller1998tensorial}
H.~G. Weller, G.~Tabor, H.~Jasak, C.~Fureby, A tensorial approach to
  computational continuum mechanics using object-oriented techniques, Comput.
  phys. 12~(6) (1998) 620--631.

\bibitem{mckay2000comparison}
M.~D. McKay, R.~J. Beckman, W.~J. Conover, A comparison of three methods for
  selecting values of input variables in the analysis of output from a computer
  code, Technometrics 42~(1) (2000) 55--61.

\bibitem{jin2005efficient}
R.~Jin, W.~Chen, A.~Sudjianto, An efficient algorithm for constructing optimal
  design of computer experiments, J. Stat. Plan. 134~(1) (2005) 268--287.

\bibitem{damblin2013numerical}
G.~Damblin, M.~Couplet, B.~Iooss, Numerical studies of space-filling designs:
  optimization of {Latin Hypercube Samples} and subprojection properties, J.
  Simul. 7~(4) (2013) 276--289.

\bibitem{kawai2014kriging}
S.~Kawai, K.~Shimoyama, Kriging-model-based uncertainty quantification in
  computational fluid dynamics, in: 32nd AIAA Applied Aerodynamics Conference,
  2014, p. 2737.

\bibitem{SMT2019}
M.~A. Bouhlel, J.~T. Hwang, N.~Bartoli, R.~Lafage, J.~Morlier, J.~R. R.~A.
  Martins, A {Python} surrogate modeling framework with derivatives, Adv. Eng.
  Softw. (2019) 102662.

\bibitem{sobester2008engineering}
A.~Sobester, A.~Forrester, A.~Keane, Engineering design via surrogate
  modelling: a practical guide, John Wiley \& Sons, (2008).

\bibitem{hastie2009elements}
T.~Hastie, R.~Tibshirani, J.~H. Friedman, J.~H. Friedman, The elements of
  statistical learning: data mining, inference, and prediction, Vol.~2,
  Springer, (2009).

\bibitem{jones1998efficient}
D.~R. Jones, M.~Schonlau, W.~J. Welch, Efficient global optimization of
  expensive black-box functions, J. Glob. Optimiz. 13~(4) (1998) 455--492.

\bibitem{mockus1978application}
J.~Mockus, V.~Tiesis, A.~Zilinskas, The application of {Bayesian} methods for
  seeking the extremum, Toward. Glob. Optimiz. 2~(117--129) (1978) 2.

\bibitem{bouhlel2019python}
M.~A. Bouhlel, J.~T. Hwang, N.~Bartoli, R.~Lafage, J.~Morlier, J.~R. Martins, A
  python surrogate modeling framework with derivatives, Advances in Engineering
  Software 135 (2019) 102662.

\bibitem{deb2002fast}
K.~Deb, A.~Pratap, S.~Agarwal, T.~Meyarivan, A fast and elitist multiobjective
  genetic algorithm: {NSGA-II}, {IEEE} Trans. Evol. Comput. 6~(2) (2002)
  182--197.

\bibitem{pinelli2010reynolds}
A.~Pinelli, M.~Uhlmann, A.~Sekimoto, G.~Kawahara, {R}eynolds number dependence
  of mean flow structure in square duct turbulence, J. Fluid Mech. 644 (2010)
  107--122.

\bibitem{mcconkey2021curated}
R.~McConkey, E.~Yee, F.-S. Lien, A curated dataset for data-driven turbulence
  modelling, Sci. Data 8~(1) (2021) 1--14.

\bibitem{moser1999direct}
R.~D. Moser, J.~Kim, N.~N. Mansour, Direct numerical simulation of turbulent
  channel flow up to re $\tau$= 590, Phys. Fluids 11~(4) (1999) 943--945.

\bibitem{lee2015direct}
M.~Lee, R.~D. Moser, Direct numerical simulation of turbulent channel flow up
  to, J. Fluid Mech. 774 (2015) 395--415.

\bibitem{vinuesa2018secondary}
R.~Vinuesa, P.~Schlatter, H.~Nagib, Secondary flow in turbulent ducts with
  increasing aspect ratio, Phys. Rev. Fluids 3~(5) (2018) 054606.

\bibitem{huser1993direct}
A.~Huser, S.~Biringen, Direct numerical simulation of turbulent flow in a
  square duct, Journal of Fluid Mechanics 257 (1993) 65--95.

\bibitem{hornshoj2021quantifying}
S.~D. Hornsh{\o}j-M{\o}ller, P.~D. Nielsen, P.~Forooghi, M.~Abkar, Quantifying
  structural uncertainties in {R}eynolds-averaged {N}avier--{S}tokes
  simulations of wind turbine wakes, Renew. Energy 164 (2021) 1550--1558.

\bibitem{emory2014visualizing}
M.~Emory, G.~Iaccarino, Visualizing turbulence anisotropy in the spatial domain
  with componentality contours, Center for Turbulence Research Annual Research
  Briefs (2014) 123--138.

\bibitem{parente2011improved}
A.~Parente, C.~Gorl{\'e}, J.~Van~Beeck, C.~Benocci, Improved k--$\varepsilon$
  model and wall function formulation for the {RANS} simulation of abl flows,
  J. Wind Eng. Ind. Aero 99~(4) (2011) 267--278.

\bibitem{menter2012explicit}
F.~Menter, A.~Garbaruk, Y.~Egorov, Explicit algebraic reynolds stress models
  for anisotropic wall-bounded flows, Progress in flight physics 3 (2012)
  89--104.

\bibitem{cherroud2023space}
S.~Cherroud, X.~Merle, P.~Cinnella, X.~Gloerfelt, Space-dependent aggregation
  of data-driven turbulence models, arXiv preprint arXiv:2306.16996 (2023).

\end{thebibliography}

\end{document}